\documentclass[12pt]{iopart}
\usepackage{epsfig}
\usepackage{graphicx}
\usepackage{iopams} 

\begin{document}
%
%
%
%
\title[Control of unstable macroscopic
]{Control of unstable macroscopic oscillations in the
dynamics of three coupled Bose condensates}

\bigskip
\author{
P Buonsante \dag\, 
R Franzosi \ddag\ and 
V Penna \dag\,
}

\address{
\dag\ 
Dipartimento di Fisica and Unit\`a C.N.I.S.M., Politecnico di Torino,
C.so Duca degli Abruzzi 24, I-10129 Torino, Italia
\\
\ddag\
Dipartimento di Fisica, Universit\`a di Firenze
\& INFN Sez. di Firenze,
Via G. Sansone 1, I-50019 Sesto Fiorentino, Italy.}

\begin{abstract}
We study the dynamical stability  of the macroscopic quantum
oscillations characterizing a system of three coupled
Bose-Einstein condensates arranged into an open-chain
geometry. The boson interaction, the hopping amplitude and
the central-well relative depth are regarded as adjustable
parameters. After deriving the {\it stability
diagrams} of the system, we identify three mechanisms to
realize the transition from an unstable to stable behavior and
analyze specific configurations that, by suitably tuning
the model parameters, give rise to macroscopic effects which are
expected to be accessible to experimental observation. 
Also, we pinpoint a system regime that realizes a
Josephson-junction-like effect.
In this regime the system configuration do not depend on the
model interaction parameters, and the population oscillation amplitude
is related to the condensate-phase difference.
This fact makes possible estimating the latter quantity, since the
measure of the oscillating amplitudes is experimentally accessible.
\end{abstract}





\section{Introduction}
\label{S:intro}

Since the first observation of Bose-Einstein condensates (BECs)
in dilute weakly interacting gases of bosons in 1995,
great efforts both experimental and theoretical have been
addressed on this subject.
Thus nowadays, thanks to the impressive progress of experimental
techniques, BECs represent a source of inspiration for new
challenging problems in quantum physics.
A notable aspect inherent in the dynamics of coupled BECs is that
the nonlinear character of their equations of motion entails a
particularly rich phenomenology where nonlinear effects including chaos
can be studied in a quantum environment.
%
Among the latter, in recent experiments have been observed
phenomena such as super-fluidity \cite{Burger_PRL86},
Josephson tunneling \cite{Cataliotti_Science293} and atom
optics \cite{Ottl_PRL95}.

If the recent past has been prolific of experiments on BECs
arranged into arrays with huge number (of the order of several hundreds)
of condensates, at the present, the experiments on chains
with a few interacting BECs (of the order of several unities) has received
a limited attention
\cite{Anker05}-\cite{Oberthaler_mt}
despite the increasing amount of theoretical work devoted to such systems.
Over the last few years, work on {\em small}-chain models has been focused
in particular on the two-well system (dimer)
\cite{Shenoy}-\cite{FP1}
and the three-well system (trimer)~\cite{Nemoto}, \cite{FP3} while their
many-well generalization has been 
studied in~\cite{Jaksch}-\cite{Kalosakas02}.
%
The theoretical prediction of macroscopic effects such as coherent atomic
tunneling~\cite{Shenoy}, population self-trapping~\cite{Raghavan},
and Josephson junction effect~\cite{Giovanazzi}, have been confirmed
by recent observations in the experimental realization 
of the dimer~\cite{Anker05}-\cite{Oberthaler_mt}.

These results stimulate the theoretic study of the BEC small chains
in order to predict new macroscopic effects that, in addition to furnishing
a deeper insight of stability properties and an increased control
of systems, open the possibility of designing significant experiments.
In this respect, the recent experimental technique discussed
in \cite{Oberthaler_bjj,Anker05} and successfully applied to realize
a two-well system provides an extremely effective tool for engineering 
small arrays. The superposition of a parabolic confining potential to a
linear optical lattice allows one to select two wells (dimer) with possibly
different depths by changing the parabola position and amplitude. Likewise,
a simple, suitable change of the latter should enable one to select 
$M$-well arrays, with $M$ arbitrary, whose (local) potentials
feature a parabolic profile.
The trimer case is thus very promising
in that, in addition to represent a practicable experimental objective,
it exhibits behaviors typical of more complex arrays.
In fact, while the simplest BEC array (the dimer)
features an integrable dynamics~\cite{Aubry}, in the trimer case
the apparently harmless addition of a further coupled condensate
is sufficient to make the system nonintegrable.
As a consequence, the trimer displays strong dynamical instabilities
in extended regions of the phase space \cite{STrimer} and a whole
new class of behaviors including chaos. In general, unstable behaviors
have been found in condensates subject to kicked pulses \cite{instGP1}
or to an anisotropic trapping potential \cite{instGP2} and within 
soliton dynamics~\cite{instGP3}.

Concerning the trimer, a conspicuous number of recent
papers has addressed various aspects of its dynamics. These include,
within a mean-field approach, the occurrence
of a transistor-like behavior \cite{Stickney}, and the self-trapping
\cite{Liu} in a triple-well asymmetric trap of repulsive bosons, the
ring-trimer dynamics with attractive interaction in the presence of a
single on-site defect \cite{Pando07}, and the chaos control through an
external laser pulse \cite{Chong}. Recently, the dynamical instability
has been studied in terms of quasimomentum modes for a $5$-well
semiclassical Bose-Hubbard model including a linear
potential \cite{KKG}.
Also, more related to purely quantum properties, second-quantized models
of the trimer have been considered to study the properties of quantum
states relevant to interwell particle exchange \cite{N2} and entanglement
\cite{Mossman}, the semiclassical quantization of the  Bogoliubov
spectrum \cite{Kol}, the structural changes of trimer eigenstates in
terms of the tunneling parameter \cite{Hiller}, and the occurrence of
vortex-like states \cite{Lee}.

In the present paper we study the classical dynamics of an asymmetric
open trimer (AOT) made of three coupled BECs arranged into an open chain
geometry, where both the interwell tunneling amplitude $T$ and the relative
depth $w$ of the central well, are regarded as adjustable parameters.
Our objective is to evidence several macroscopic effects
that can take place in the AOT dynamics to provide a useful guide for
future experiments. 
In section~\ref{S:Dyn} we introduce the semiclassical equations of the trimer
mean-field dynamics. In section~\ref{S:FP}, after determining the solutions
relevant to the fixed points of motion equations, i. e. the proper modes of
trimer dynamics, we focus mostly on establishing, via standard procedures,
their stability character for experimentally significant values of effective
parameters $\tau := T/UN$ and $\nu := w/UN$, where $U$ and $N$ are the
effective interatomic scattering and the total boson number, respectively.
In section~\ref{S:SD} we review the (three) {\it stability diagrams}
(one for each class of fixed points) that summarize characters and properties
of fixed points. These supply an exhaustive, operationally valuable, account
of the system stability properties and allow to determine the stable/unstable
charater of the system evolution for initial conditions situated, in the phase
space, close to a given fixed point.
Both the derivation and the use of such diagrams, rapidly sketched
in \cite{PFLet1}, are discussed here (and in the relevant appendices)
with some detail together with various new applications. 

By using the {\it stability diagrams} as a map for detecting the critical
behaviors of trimer, in section~\ref{S:macroeff} we find parameter-tuned macroscopic effects (both in the dimeric and nondimeric regime) and evidence
some features that may prove experimentally relevant. Such effects are shown
to arise when crossing either the boundaries of regions with different
stability characters or the boundaries of forbidden regions or, finally,
by exploiting the coalescence/bifurcation mechanism characterizing certain
fixed points.
In section~\ref{S:CDWI} we consider the special case of $\pi$-like states
relevant to the fixed points of the regime charaterized by a central
depleted well. We investigate the unexpected relation between the
average population oscillations and the initial phase difference of the 
lateral condensates, suggesting a possible experiment. 
Finally, section~\ref{S:conc} is devoted to make some concluding remarks.

\section{Dynamical equations of the AOT and fixed points}
\label{S:Dyn}

The quantum Hamiltonian of the AOT on a linear chain
reads~\cite{Jaksch}
\begin{equation}
\label{E:Ham3}
H_3 = 
\sum^{3}_{i=1} (U n_i^2 -v n_i) -w n_2 
\!-\!\frac{_T}{^2} \left [ a^{\dagger}_2(a_{1}+a_{3}) \!+\! \rm{H.C.} \right ] 
\end{equation}
where site boson operators $a_i^{\dagger},a_i$ satisfy standard
commutators $[ a_{i} ,a^{\dagger}_k] = \delta_{ik}$, $n_i = a_i^+ a_i$, 
$v$ is the energy offset of the wells, $w$ is the central-well relative
depth, $T$ is the hopping amplitude and $U \simeq 2\pi a \hbar^2/m$ 
($a$ the scattering length, $m$ the boson mass) is the on-site boson interaction.
The total boson number $N = \sum_i n_i$, commutes with Hamiltonian
(\ref{E:Ham3}). 
The exact form of parameters $U$ and $T$ is calculated in~\cite{gerb}
showing how such quantities depend on $m$, on the laser
wavelength $\lambda$, and on the potential depth $V_0$ of potential
$V_{op}(x)= V_0 \sin^2(2\pi x/\lambda)$ determining the optical lattice.
The 3-well linear system, obtained by superposing the optical potential
to parabolic potential $V_p(x)= m \omega^2 x^2/2$ confining
bosons~\cite{Anker05}, \cite{Oberthaler_bjj}, is expected to have 
lateral wells with depth $v$ smaller then the central-well depth
$v+w$ due to the presence of the parabolic potential.   
The dynamics of the AOT with strong tunnelling amplitudes
($T/U>>1$) and large average numbers of atoms per well can be
described by the mean-field form of (\ref{E:Ham3}).
The method developed in \cite{Amico2} combines the coherent-state picture
of quantum systems with the time-dependent variational approach~\cite{Zhang}
providing the semiclassical version of quantum model (\ref{E:Ham3})
in which the dynamical canonical variables are the expectation values 
$z_i := \langle Z| a_i |Z \rangle$, $z^*_i := \langle Z| a^+_i |Z \rangle$.
The so-called trial macroscopic wave function $|Z \rangle$ is standardly
assumed to be  $|Z \rangle = \prod_{i} |z_i \rangle$, where $1 \le i\le 3$
and $|z_j \rangle$ are the Glauber coherent states such that
$a_i |z_i \rangle = z_i \, | z_i\rangle$ for each $i$.
Then, the resulting mean-field Hamiltonian reads
\begin{equation}
\label{E:H3}
{\cal H}_3 = \sum_{j=1}^3\left[U |z_j|^4 - v |z_j|^2 \right] 
- w |z_2|^2-\frac{_T}{^2}[\, z_2^* (z_1+z_3) +{\rm C.C.}] \, ,
\end{equation}
which, through the Poisson brackets $\{z_j^*,z_k\}=\frac{i}{\hbar}\delta_{j k}$,
yields the dynamical equations
\begin{equation}
\label{E:dyneq}
\left\{
\begin{array}{ll}
i \hbar \dot z_j \!=\! \left(2U|z_j|^2-v\right)z_j
-\frac{T}{2}z_2 & j=1,3\\
i \hbar \dot z_2\!=\! \left(2U|z_2|^2-v\! -\! w\right)z_2 
\!-\!\frac{T}{2}\left(z_1\!+\!z_3\right) &
\end{array}
\right.
\end{equation}
The validity of such a semiclassical picture is confirmed 
experimentally~\cite{Anker05}, \cite{Oberthaler_bjj} for the
time scale (hundreds of milliseconds) of the
macroscopic oscillations in a two-well system. 
The effect of approximating the system state $|Z \rangle$
through a product of localized (coherent) states $\prod_i |z_i \rangle$
appears on a much longer time scale~\cite{RSK} when $|Z \rangle$
must be expressed as a superposition of Fock states 
$|{\vec n} \rangle= |n_1, n_2, n_3 \rangle$ 
($a^+_i a_i |{\vec n} \rangle = n_i |{\vec n} \rangle$) 
to include (quantum) delocalization effects.
An early analysis of equations (\ref{E:dyneq}) for $w=0$ is reported in 
\cite{Finlayson}, \cite{Henn}.
It is easy matter to check that system (\ref{E:dyneq}) embodies
the conservation of boson total number $N=\Sigma_j |z_j|^2$,
namely $\{N, H \}=0$. Constant $N$ can be exploited to introduce
new rescaled variables 
$Z_j \equiv z_j e^{-i v t/ \hbar} /{\sqrt N} $
whose evolution is governed by
\begin{equation}
\label{E:dyneq2}
\left\{
\begin{array}{ll}
%
iZ'_j \!=\! 2|Z_j|^2 \, Z_j-\frac{\tau}{2}Z_2 
& 
j=1,3
\\
iZ'_2 \!=\! \left(2|Z_2|^2 - \nu \right)Z_2
-\frac{\tau}{2}\left(Z_1\!+\!Z_3\right) &
\end{array}
\right.
\end{equation}
where
\begin{equation}
\label{E:rescP}
\quad \nu ={w}/{(U N)},
\quad \tau ={T}/{(U N)}, \quad {\tilde t} ={U N t}/{\hbar} \, ,
\end{equation}
and the prime denotes derivation with respect to the rescaled 
time variable $\tilde t$. This shows that the two significant parameters
of the trimer dynamics are $\nu$ and $\tau$ (namely, at fixed $N$, $w/U$
and $T/U$), parameter $v$ being incorporated in the time dependent factor
$e^{-i v t/ \hbar}$ of new variables $Z_j$.
%
%
Furthermore, it is easy to verify that the condition
$z_1 (t)= z_3 (t)$ is valid for $t>0$
(or, equivalently, that $Z_1(\tilde t)=Z_3(\tilde t)$ holds 
for every $\tilde t>0$) if, initially, $z_1 (0)= z_3 (0)$.
Such a condition identifies an integrable subregime 
where the lateral condensates have the same population and phase
(see, e. g., \cite{Finlayson}, \cite{Andersen93a} and \cite{Lphys3})
which in the following we shall refer to as {\it dimeric regime}.


%
\subsection{Fixed points}
\label{S:FP}
We can get insight about the behaviour of the system in different
dynamical regimes through the study of special exact solutions along
with their stability character. With the latter we mean the simplest
(single mode) periodic orbits in the phase space of the form
$Z_j(t') = x_j e^{im t'}$. Substituting the latter in equations
(\ref{E:dyneq2}) provides the equations for variables $x_j$.
Such equations can be equivalently obtained from
$i {Z'}_j = 0= \{Z_j, {\tilde H} -m {\tilde N} \}$ 
with ${\tilde H} = {\cal H}_3 (Z_i , Z^*_i)$,
where the chemical potential $m$, a Lagrange multiplier, 
allows one to include 
the conserved quantity ${\tilde N} = \sum_i |Z_i|^2 =1$
derived from the total boson number $N= \sum_i |z_i|^2$.
Owing to the condition ${Z'}_i = 0$ we name fixed-point equations the
equations determining $x_i$, $i = 1,2,3$.
%
%
These have the form
\begin{equation}
\label{E:fpsys}	
0= \left(2 \,x_j^2-m\right)\,x_j-\frac{\tau}{2}\,x_2\, ,\,\,\,
0= \left(2\,x_2^2-m - \nu\right)\,x_2 -\frac{\tau}{2}\,\left(x_1+x_3\right)
\end{equation}
where $j=1,3$ and the fixed-point coordinates ${x_j}$ 
are normalized to unity: $\sum_j x_j^2=1$. In equation \ref{E:fpsys}
we have exploited the system symmetry, inherited from equations \ref{E:dyneq2},
under a global phase shift $x_j \mapsto x_j \exp(i \Phi)$ to restrict coordinates ${x_j}$ (which are in principle complex numbers) to the
real axis. Let us summarize the solutions of system (\ref{E:fpsys}).
\medskip

\noindent
{\it Central depleted well}. This parameter-independent
solution is obtained when the central well is
strictly depleted, that is $x_2=0$, whereas the lateral ones
have the same population and opposite phases.
Due to the constraint on the number it must necessarily be 
$x_1=-x_3 = \pm \sqrt{m/2}=\pm 1/\sqrt 2$, $m=1$. In summary
\begin{equation}
x_1=-x_3 = \pm 1/\sqrt 2 \, , \quad  x_2 = 0 \, .
\label{cdwfp}
\end{equation}
We shall
refer to this solution as {\it central depleted well} (CDW). 
%
Since configurations where either one or both of the lateral condensates
have zero population are forbidden by equations~(\ref{E:fpsys}), we are
left with the following two possibilities.
\medskip

\noindent
{\it Dimeric solutions}.
Similar to (\ref{E:dyneq}) and (\ref{E:dyneq2}), equations (\ref{E:fpsys})
possess dimeric solutions characterized by $x_2\neq0$ and $x_1=x_3 \neq 0$.
These solutions can be expressed via the polar representation
\begin{equation}
x_1=x_3= {\cos \theta}/{\sqrt 2}  \, , \qquad x_2= \sin \theta \, .
\label{dfp}
\end{equation}
Introducing this parametrization in (\ref{E:fpsys}) one finds that
the tangent $\tan \theta=:{\alpha}$ of the angular coordinate
$\theta$ identifying each fixed point is given by the real roots of
\begin{equation}
{\cal P}({\alpha};\tau,\nu)\equiv {\alpha}^4+
\frac{A_{\nu}}{\tau} {\alpha}^3- \frac{B_{\nu}}{\tau}{\alpha}-1 =0 \, ,
\label{dfp_c}
\end{equation}
with $A_{\nu} =\sqrt 2(2-\nu)$, $B_{\nu} =\sqrt 2(1+\nu)$.
In the following we shall refer to this class of solutions
as {\it dimeric fixed points} (DFP) whose derivation is
discussed in \ref{S:DR}.
\medskip

\noindent
{\it Non-dimeric solutions}.
The remaining single-mode solutions of system (\ref{E:fpsys}),
characterized by $x_1\neq x_3$, $x_j\neq0$, will be referred to as 
{\it non-dimeric fixed points} (NFP). In this case it is appropriate
to switch to the alternative set of coordinates 
$X_1= (x_1+x_3)/{\sqrt 2}$, $X_2=x_2$, and  $X_3=(x_1-x_3)/{\sqrt 2}$
before setting
\begin{equation}
\label{ndfp_p}
X_1= R \cos \theta,\quad  X_2= R  \sin \theta \, .
\end{equation}
In this case the radial coordinate $R$ reads
\begin{equation}
\label{ndfp_ra}
R({\alpha};\,\tau)=\sqrt{\frac{(1+\alpha^2)(2\sqrt 2-\alpha\, 
\tau)}{2\sqrt 2 (2+\alpha^2)}}
\, , \quad
{\alpha}=\tan \theta := \frac{X_2}{X_1}
\end{equation}
and depends on the Hamiltonian parameter $\tau$ and on parameter $\alpha$.
By introducing this parametrization in system (\ref{E:fpsys}) one finds that
$\alpha$ is given by the real roots of
\begin{equation}
\label{ndfp_pol}
{\cal P}({\alpha};\tau,\nu)\equiv {\alpha}^4+
\frac{A_{\nu}}{\tau} {\alpha}^3-
\frac{B_{\nu}}{\tau} {\alpha}-\frac{4}{3}=0 \, ,
\end{equation}
where $A_{\nu}={2 \sqrt 2(2-\nu)}/{3}$, $B_{\nu}={4\sqrt 2(1+\nu)}/{3}$.
Since the physical solutions of system (\ref{E:fpsys}) have to be real it
is clear that the roots which make $R({\alpha})$ or
\begin{equation}
\label{ndfp_last}
X_3(\alpha)\!=\!\pm\sqrt{1-X_1^2-x_2^2}\!=\!
\pm\sqrt{\frac{4\!+\!\sqrt 2 \alpha(1+\tau \alpha^2)}{4(2+{\alpha}^2)}}
\end{equation}
imaginary, must be discarded.
The next section contains a qualitative discussion of the three stability
diagrams describing the entire set of solutions of system (\ref{E:fpsys})
for any choice of $\tau$ and $\nu$. A detailed derivation of such fixed
points is given in \ref{S:NR}.


\section{Stability diagrams}
\label{S:SD}

\begin{figure}[htpb]
\centering \epsfig{figure=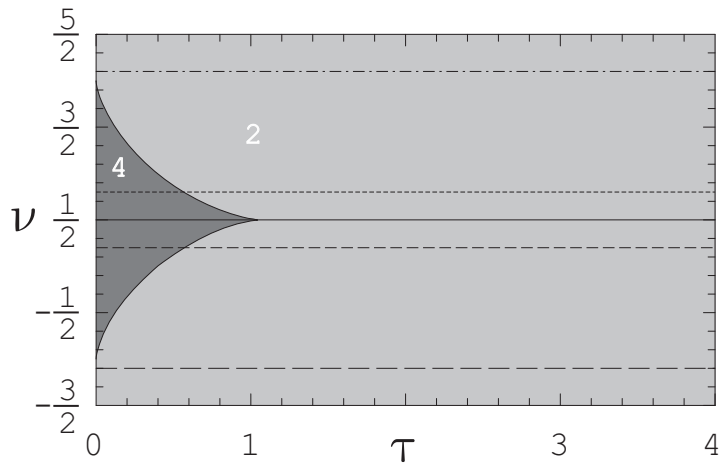,width=0.6\textwidth}
\caption{
Number of dimeric fixed points 
on varying parameters $\tau$ and $\nu$. The border 
of the lobe where four solutions are found is 
described by (\ref{E:Dlobe}).}
\label{fig:DN}
\end{figure}

\begin{figure}[htpb]
\centering \epsfig{figure=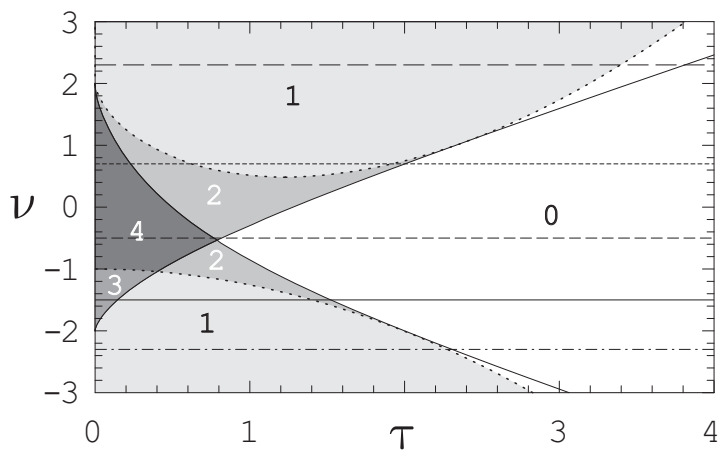,width=0.6\textwidth}
\caption{
Number of non-dimeric fixed points on varying parameters $\tau$ and $\nu$. 
The same shade of grey fills regions characterized 
by the same number of fixed points (shown).}
\label{fig:NN}
\end{figure}

The stability diagrams introduced in \cite{PFLet1} allow one to know both
number and stability character of the fixed points of trimer dynamics for
any choice of significant parameters $\tau$ and $\nu$. In the previous
section we have shown that vector ${\bf x} = (x_1, x_2, x_3)$ describing
a fixed point can be parametrized by the angular coordinate $\theta$ of
the two-dimensional polar representations (\ref{dfp}), and (\ref{ndfp_p})
and (\ref{ndfp_ra}) for the DFP and the NFP, respectively.
To determine parameter $\theta$ (and thus the vector ${\bf x}$ relevant to
a given fixed point) we must find the roots of the fourth-degree polynomial
\begin{equation}
{\cal P}(\alpha;\tau,\nu) = {\alpha}^4+\frac{A_\nu}{\tau} 
{\alpha}^3+\frac{B_\nu}{\tau} {\alpha}+C = 0
\label{GenPol}
\end{equation}
in the unknown quantity $\alpha =\tan \theta$. This allows one to determine
$\alpha = x_2/\sqrt{2} x_1$ in the dimeric case with $C=-1$
(see (\ref{dfp_c})), and 
$\alpha = x_2/\sqrt{2}(x_1+x_3)$ in the non-dimeric case with $C=-4/3$
(see (\ref{ndfp_pol})). In the latter case vector $\bf x$ is obtained by
using the constraint $x_1^2+x_2^2+x_3^2=1$.
Therefore, for a given value of parameters $\tau$ and $\nu$, the number of
roots of (\ref{GenPol}) establishes how many DFP and NFP there exist.
This information is summarized in figures \ref{fig:DN} and \ref{fig:NN}.

\begin{figure}[htpb]
\centering \epsfig{figure=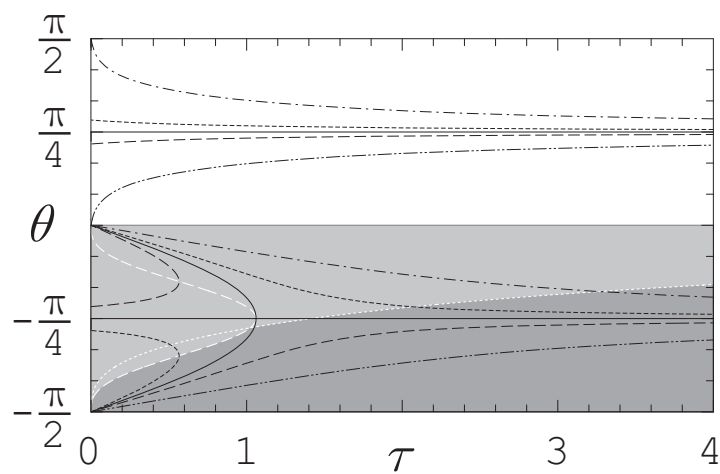,width=0.6\textwidth}
\caption{
Stability diagram for the dimeric fixed points (DFP).
The fixed points belonging to the white, medium grey and
dark grey  regions are always minima, unstable saddles and
maxima, respectively. The latter two regions are divided
by the curve $\tau=d(\tan \theta)$ defined in \ref{S:DR}.
The tip of the ``parabolic'' branch moves along the white dashed curve.
The white dotted and dashed curves are $d_1(\tan \theta)$
and $d_2(\tan \theta)$, respectively (see \ref{S:DR} for details).
Note that the curves drawn with diffrent dashing styles, correspond to
the five values of $\nu$ selecting the straight horizontal
lines reported with the same dashing styles in figure~\ref{fig:DN}.
}
\label{fig:DS}
\end{figure}

\begin{figure}[htpb]
\centering \epsfig{figure=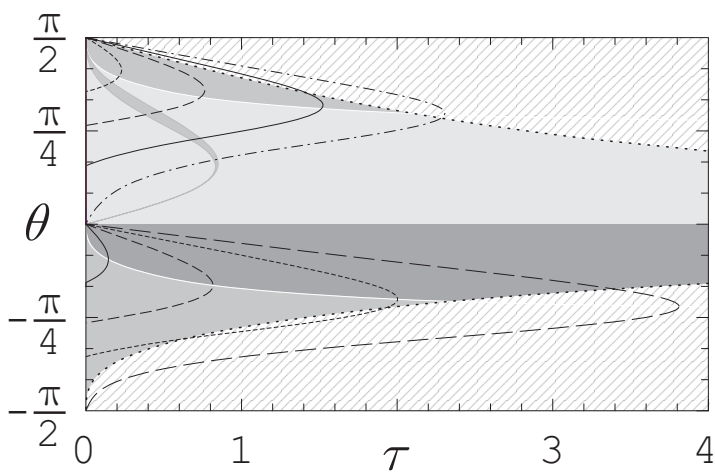,width=0.6\textwidth}
\caption{
Stability diagram for the non-dimeric fixed points (NFP). The fixed points 
belonging to the light, medium  and dark grey regions are always stable
saddles, unstable saddles and maxima, respectively. Note that the white
solid curves are the loci of the tips of the two parabolic branches. 
The patterned regions must be ignored, since the solutions 
found there yield complex $x_j$'s 
(see \ref{S:NR} for details).
The curves drawn with diffrent dashing styles, correspond to
the five values of $\nu$ selecting the straight horizontal
lines reported with the same dashing styles in figure~\ref{fig:NN}.}
\label{fig:NS}
\end{figure}

The stability character of the fixed points solving 
equations (\ref{E:fpsys}) is depicted in the stability diagrams 
of figures~\ref{fig:DS} and \ref{fig:NS}
for the DFP and the NFP case, respectively. In such figures any point in
the plane $\theta-\tau$ is implicitly associated to a definite value of
parameter $\nu$, owing to equation (\ref{GenPol}).
This can be seen by rewriting (\ref{GenPol}) as
\begin{equation}
\label{E:GtauC}
\tau[\tan(\theta);\nu] 
=- \frac{\left[A_\nu \tan^2(\theta)+B_\nu\right]\tan(\theta)}{\tan^4(\theta)+C} \, ,
\end{equation}
which, for a given value $\nu = \tilde \nu$, defines 
a curve $\tau[\tan(\theta);\tilde \nu]$ in the plane $\theta$-$\tau$
consisting of two or more non-connected branches, 
both in the dimeric ($C=-1$) and the non-dimeric case ($C=-4/3$).
At the operational level, in both figures, for a given choice
$\tau \equiv \tilde \tau$ and  $\nu \equiv \tilde \nu$, configuration
parameter $\theta$ can be obtained by intersecting the curves
$\tau[\tan(\theta);\tilde \nu]$, for the dimeric and non-dimeric case,
and the straight line $\tau=\tilde \tau$.
Once (triplet $\tau$, $\nu$, $\theta$ of) a fixed point has been
identified, its stable/unstable character can be evinced from figures
\ref{fig:DS} or \ref{fig:NS} thanks to the colour corresponding to a
given value of $\theta$ and $\tau$ (see the relevant figure captions).
The linear stability analysis determining the character of fixed points
(and, in practice, the whole structure of such phase diagrams) is
discussed in \ref{S:ASD}.
%
%
%
%
As an example, in figures~\ref{fig:DN} and \ref{fig:NN} are
reported with different dashing styles, five straight horizontal
lines corresponding to several values of the $\nu$.
The five curves drawn in figures~\ref{fig:DS} and \ref{fig:NS},
refer to the straight lines drawn with the same dashing styles
in figures~\ref{fig:DN} and \ref{fig:NN}, for the dimeric and
non-dimeric case, respectively.
%
%
Notice that in general these curves consist of two or 
more non-connected branches. In the dimeric case there 
are always two unbounded branches featuring an asymptote 
at $\theta=+\pi/4$ or at $\theta=-\pm \pi/4$. For $\nu \in [-1,2]$ 
a further, roughly parabolic, bounded branch appears. 
This means that for any choice of the parameters $\tau$ 
and $\nu$ there are always at least two DFPs, and that, 
for suitable choices of these parameters, 
two more fixed points appear. 

In the non-dimeric case, figure~\ref{fig:NS}, the curves  
$\tau[\tan(\theta);\nu]$ consist of either one or two 
bounded parabola-like branches, hence featuring zero, 
two or four intersection with a constant-$\tau$ straight 
line. However, for certain values of the parameters, not 
all of these intersections represent fixed points, i.e. 
real solutions of system (\ref{E:fpsys}). 
Indeed, as it is discussed in \ref{S:NR}, some of 
the real roots of polynomial ${\cal P}(\alpha;\tau,\nu)$
--more precisely the ones lying in the patterned regions of
figure~\ref{fig:NS}-- result in complex-valued solutions of
system (\ref{E:fpsys}) and therefore must be discarded. 
In particular, the patterned (forbidden) regions of figure~\ref{fig:NS}
can be shown to correspond to the white region in the $\tau$-$\nu$ plane
of figure~\ref{fig:NN} where no solution is permitted.
Figures~\ref{fig:DS} and \ref{fig:NS} also supply important information
concerning fixed points and their characteristic energies. 
Referring to the different colours
used in such figures, a fixed point is a (stable) minimum, a stable saddle, 
an unstable saddle or a (stable) maximum of the energy function depending on 
whether the region it lies in is white, light grey, 
medium grey or dark grey. 

\begin{figure}[htpb]
\centering \epsfig{figure=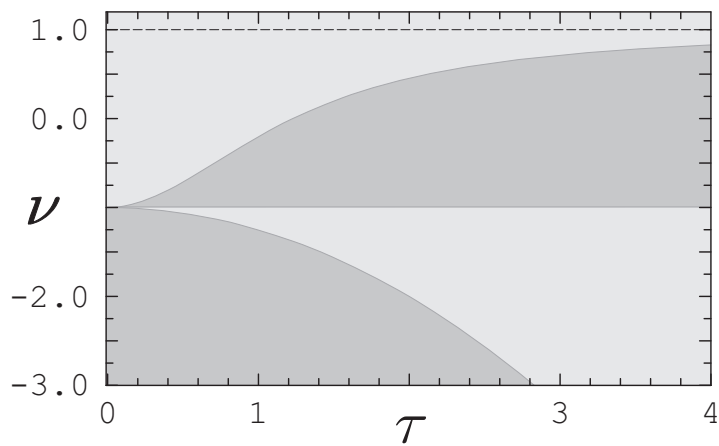,width=0.6\textwidth}
\caption{ 
Stability diagram for the CDW 
saddle point. Note that large stability regions appear 
(light grey filling). The boundaries are the straight line $\nu =0$ and
the  curves $\tau=c_1(\nu)$ and $\tau=c_2(\nu)$, described by (\ref{E:cdwB}).
The asymptote $\nu=1$ of the latter curve is 
also displayed (dashed line).}
\label{fig:CDW}
\end{figure}

As we mentioned above, other than the DFPs and the NFPs, a further
fixed point, the CDW configuration, is present for any value of
$\tau$ and $\nu$.
The parameter independence of the relevant configuration
${\bf x}=(1/\sqrt 2,0,-1/\sqrt 2)$, 
makes the use of a polar representation unnecessary, 
and allows one to draw its stability diagram in the parameter 
plane $\tau$-$\nu$ in a direct way. This is illustrated in 
figure~\ref{fig:CDW}.

\section{Macroscopic effects in the AOT dynamics}
\label{S:macroeff}

A careful look at stability diagrams \ref{fig:DS} and \ref{fig:NS},
in addition to provide, at first glance,
number and properties of the fixed points of AOT dynamics
for any choice of $\tau$ and $\nu$, allows one
to single out at least three ways to produce macroscopic, and hence
experimentally observable, effects by means of a simple change of
model parameters.

{\it Crossing boundaries of regions with a different character.}
For suitable choices of $\nu$, curves  
$\tau[\tan(\theta);\nu]$ may cross the boundary dividing
regions with different stability properties.
Thus, by slightly and adiabatically changing
the Hamiltonian parameters (e.g. the optical potential
amplitude) the system can modify significantly its behavior.
This is the case of (one of) the asymptotic branches of the dimeric 
curves, which, for any $\nu > 2\sqrt{14}-7\approx 0.48$, crosses 
the boundary between the maxima and the unstable saddles 
region (see, e. g., the branch in figure \ref{fig:DN} corresponding 
to $\nu = 0.6$ in figure \ref{fig:NN}).  Likewise, in the non-dimeric case,
one of the parabola-like branches of curves $\tau[\tan(\theta);\nu]$
always crosses the interesting unstable ``isthmus'' 
splitting the large stable (light-grey) region 
in figure~\ref{fig:NS}, provided that $\nu<2$. 

{\it Coalescence}/{\it bifurcation of fixed points.}
At the ``tips'' of the parabola-like branches of curves
$\tau[\tan(\theta);\nu]$, a {\it coalescence}/{\it bifurcation} 
{\it effect} of pairs of fixed points takes place.
Hence, for some values $\tilde \tau$ of the
hopping parameter it is possible to find a value $\tilde \nu$ 
with two coincident solutions of fixed-point equations located
at the parabolic-branch tip. 
By varying $\nu$, the branch tip moves away from $\tilde \tau$,
and the coincident solutions either split (i.e. bifurcate) or
disappear accordingly. Let us emphasize that in the non-dimeric
case (figure~\ref{fig:NS}) the coalescence phenomena always involve 
fixed points with different stability characters, whereas, in the 
dimeric case (figure~\ref{fig:DS}) only the coalescence of two 
fixed points of the same kind may happen, provided that $\nu >
2\sqrt{14} - 7$ (see next section).

{\it Crossing forbidden-region boundaries.}
Significant changes in the phase-space structure of the system 
are expected as well where the non-dimeric curves
$\tau[\tan(\theta);\nu]$ cross the boundaries of forbidden zones. 
Indeed, in this case, a slight change in the parameters
may cause the sudden appearance/disappearance of a fixed point.
As to the CDW fixed points, which --according to the Hessian 
signature-- is always a saddle, figure~\ref{fig:CDW} clearly 
shows that it may display an either stable or unstable 
character, depending on the choice of parameters 
$\tau$ and $\nu$ 
%
%

Based on these three mechanisms,
we discuss the results of several numeric 
simulations evidencing how macroscopic effects are induced 
in the AOT dynamics by slight changes in parameters $\tau$ and $\nu$
(similar effects on the stability 
of the many-well system with $U<0$,
of vortexlike modes in a 1D array and of localized modes in a 2D array
have been studied in
~\cite{Joha}, 
\cite{Paraoanu} and \cite{Kalosakas02}, respectively).
These entail that trajectories based on the same initial configuration may
exhibit an either regular or chaotic behavior. 
The character of the AOT dynamics, can be conveniently investigated
by constructing appropriate Poincar\'e sections
within the four-dimensional {\it reduced} phase-space spanned by 
the ({\it reduced}) set of canonical variables 
${\phi_1, \xi_1,\phi_2, \xi_2}$. 
This reduction is obtained via the canonical transformation $(z_i, z^*_i)$
$\to (\phi_1,\phi_2, \psi, \xi_1,\xi_2, N)$, where
\begin{equation}
\label{E:redvar}
\begin{array}{ll}
\phi_1=\theta_2-\theta_1, & \xi_1 = (|z_2|^2+|z_3|^2-|z_1|^2)/N,\\
\phi_2=\theta_3-\theta_2, & \xi_2 = (|z_3|^2-|z_2|^2-|z_1|^2)/N,\\
\psi=\theta_3+\theta_1,  & N = (|z_1|^2+|z_2|^2+|z_3|^2),\\
\end{array} 
\end{equation}
once the condition $N= {\rm const}$ is taken into account and one
notices that the $\psi$ evolution is nonautonomous
(see \cite{STrimer} for details on
the reduced dynamics).

We have numerically integrated the trimer equations of motion in the
reduced coordinates, by a first-order bilateral symplectic algorithm.
Moreover, we have qualitatively investigated the degree of regularity
of the ensuing trajectories
by means of Poincar\'e sections. The latter are obtained
by projecting the points
belonging to a given submanifold, characterized by a given 
value of one of the four reduced variables,  onto the plane 
spanned by two of the three remaining variables. 
All of the Poincar\'e sections reported in the following
refer to a set of trajectories based on an array 
of starting points surrounding a significant configuration 
of the system, usually a fixed point, of reduced coordinates 
$(\tilde \phi_1,\tilde \xi_1,\tilde\phi_2, \tilde \xi_2)$. 
In every instance the points belonging to the submanifold 
$\xi_2=\tilde \xi_2$ have been projected onto the 
$\phi_1$-$\xi_1$ plane.

\begin{figure}[htpb]
\centering \epsfig{figure=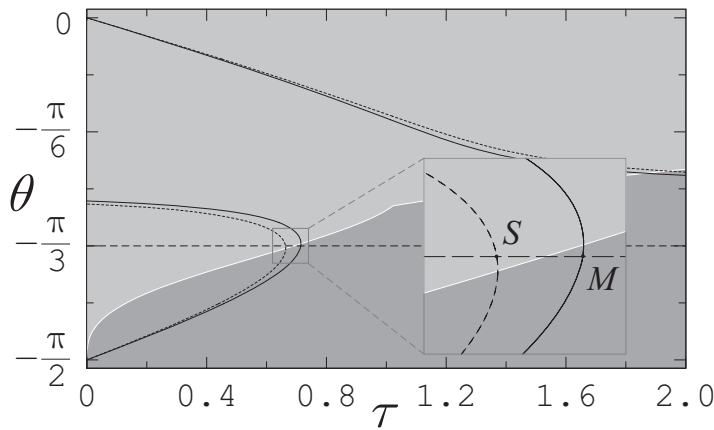,width=0.6\textwidth}
\caption{
Parameter-induced macroscopic change 
in the stability of the trimer dynamics. The figure shows a portion of the lower half plane of dimeric stability diagram \ref{fig:DS}, along with the dimeric $\tau$-curves (see~(\ref{E:GtauC}), (\ref{E:Dpc})) relevant to $\nu=\nu_{\rm M}=2/3$ (solid) and $\nu=\nu_{\rm S}=17/24$ (dashed). The white dotted and dashed curves appearing in figure~\ref{fig:DS} have been omitted for clarity. From the figure it is clear that, for both of the considered values of $\nu$, there exists a suitable value of the hopping parameter ---$\tau_{\rm M}=7/4\sqrt 6$ and $\tau_{\rm S}=13/8 \sqrt 6$, respectively --- producing a solution at $\theta=-\pi/3$ (horizontal dashed line). Hence slight differences in the parameters may result in different stability characters for the same configuration ($\theta=-\pi/3$). Indeed M is a (stable) maximum, whereas $S$ is an unstable saddle.}
\label{fig:Dexp}
\end{figure}

%
%
\subsection{Priming instability in breather-like regular oscillations}
\label{S:priming}
The first macroscopic effect we discuss here has been briefly discussed
in \cite{PFLet1} and concerns the dimeric configuration
${\bf x}=(1/\sqrt 8,-\sqrt 3/2,1/\sqrt 8)^{\rm t}$, 
--with $N/8$ bosons in each lateral well and the remaining
$3N/4$ in the central one-- which is represented by the angular variable,
$\theta=-\pi/3$. One can show that the latter is a (dimeric) fixed point
provided that $4\sqrt 6\, \nu+8 \,\tau-5\sqrt 6=0$. Furthermore, such a
fixed point is either a maximum or an unstable saddle depending on whether
$\nu$ is smaller or larger than the critical value $9 \sqrt 6/32$. 
This is clearly visible in figure~\ref{fig:Dexp}, where two different
parameter choices $(\tau_{\rm M},\nu_{\rm M} )$ and 
$(\tau_{\rm S},\nu_{\rm S} )$, shown with the letters $M$ and $S$,
are considered. 

The crossing of the boundary separating the stable and unstable
regions has evident consequences. The Poincar\'e sections (reported in
\cite{PFLet1}) of trajectories, based on a set of points closely
surrounding the fixed point under concern
$(\phi_1, \xi_1,\phi_2, \xi_2)=(\pi,3/4,-\pi,-3/4)$,
mirror the dramatic change in the dynamical behaviour involved in such
a slight parameter change. While choice $(\tau_{\rm M},\nu_{\rm M} )$
results in the neat and regular Poincar\'e sections characterizing a
stable fixed point, with choice $(\tau_{\rm S},\nu_{\rm S} )$ sections
densely fill an area considerably larger than the previous example, 
despite the relevant trajectories are based on the very same points  
\cite{PFLet1}.

\subsection{Loss of anti-breather coherence in the {\it isthmus} domain}
\label{S:losscoher}
\begin{figure}[htpb]
\centering 
\epsfig{figure=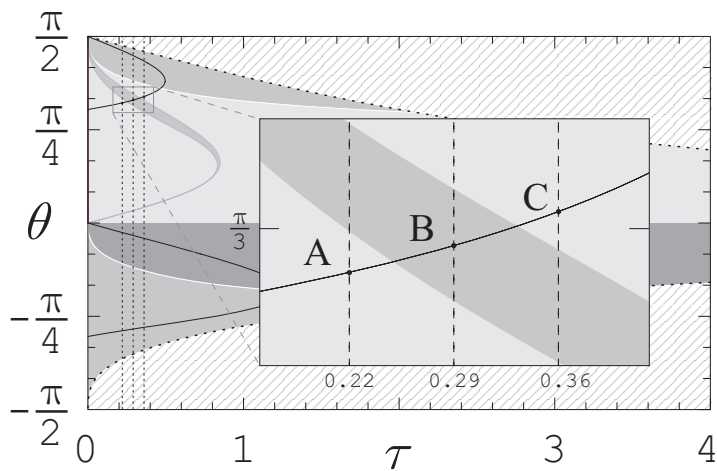,width=0.6\textwidth}
\caption{
Parameter-induced macroscopic 
change in the stability of the trimer dynamics. The figure shows the
magnification of a portion of the upper half plane of the non-dimeric
stability diagram, figure~\ref{fig:NS}, along with the non-dimeric
$\tau$-curve
(see~(\ref{E:GtauC}), (\ref{E:Gpc})) relevant to $\nu=0$ (solid) and the
straight lines corresponding to three different choices of the hopping
parameter, $\tau_{\rm A}$, $\tau_{\rm B}$ and $\tau_{\rm C}$ (dashed).
The ensuing intersections are denoted A, B and C, respectively. Notice
that B belongs to the narrow unstable region (which we term {\it isthmus}),
whereas A and C belong to the wide stable region surrounding it. The
changes in the stability character of the fixed point resulting from
such a relatively slight increase of $\tau$ are evident in
figure~\ref{fig:Nexp}.}
\label{fig:Gexp}
\end{figure}

\begin{figure*}
\centering 
\begin{tabular}{ccc}
\epsfig{figure=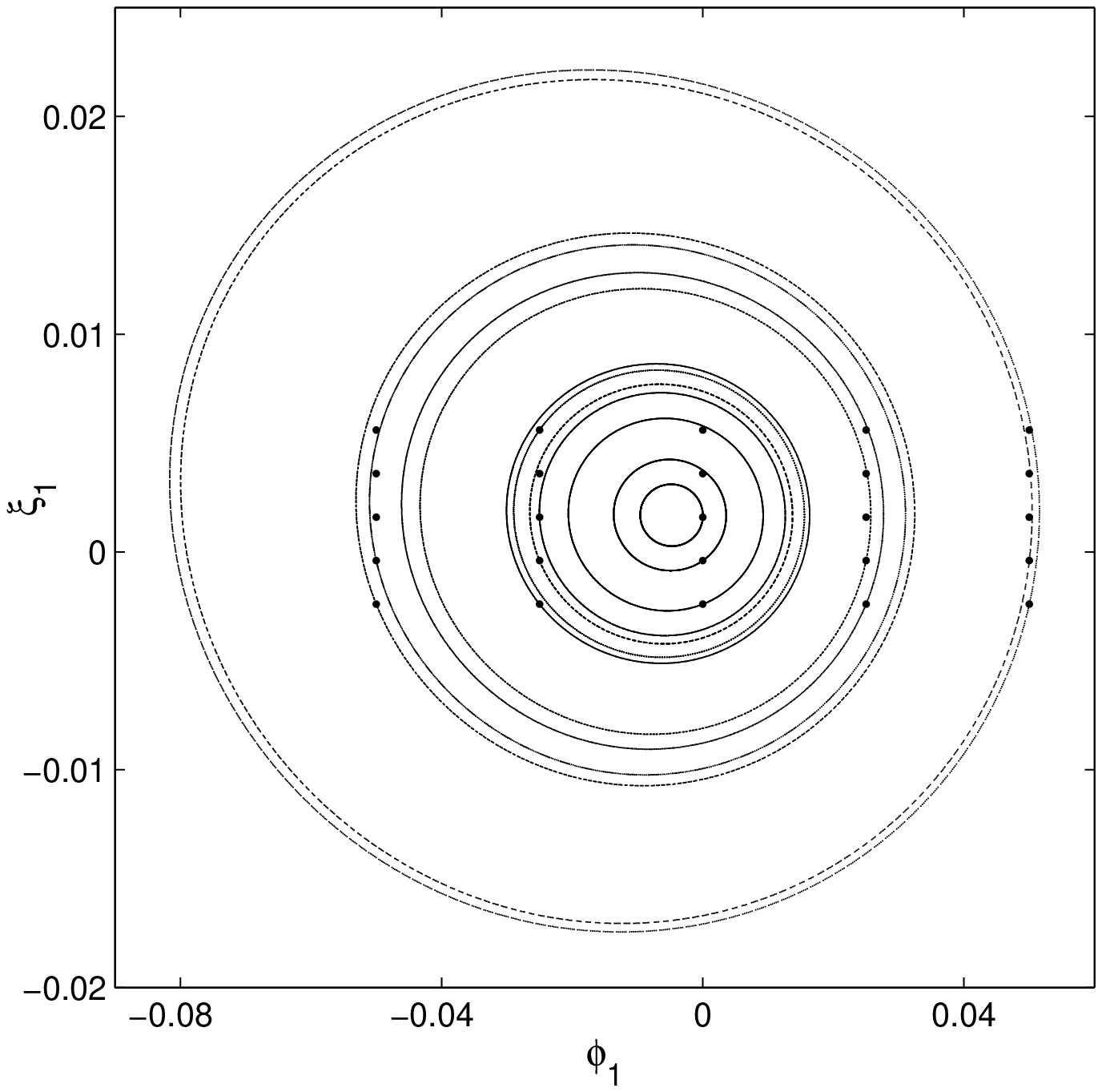,width=0.3\textwidth}
\epsfig{figure=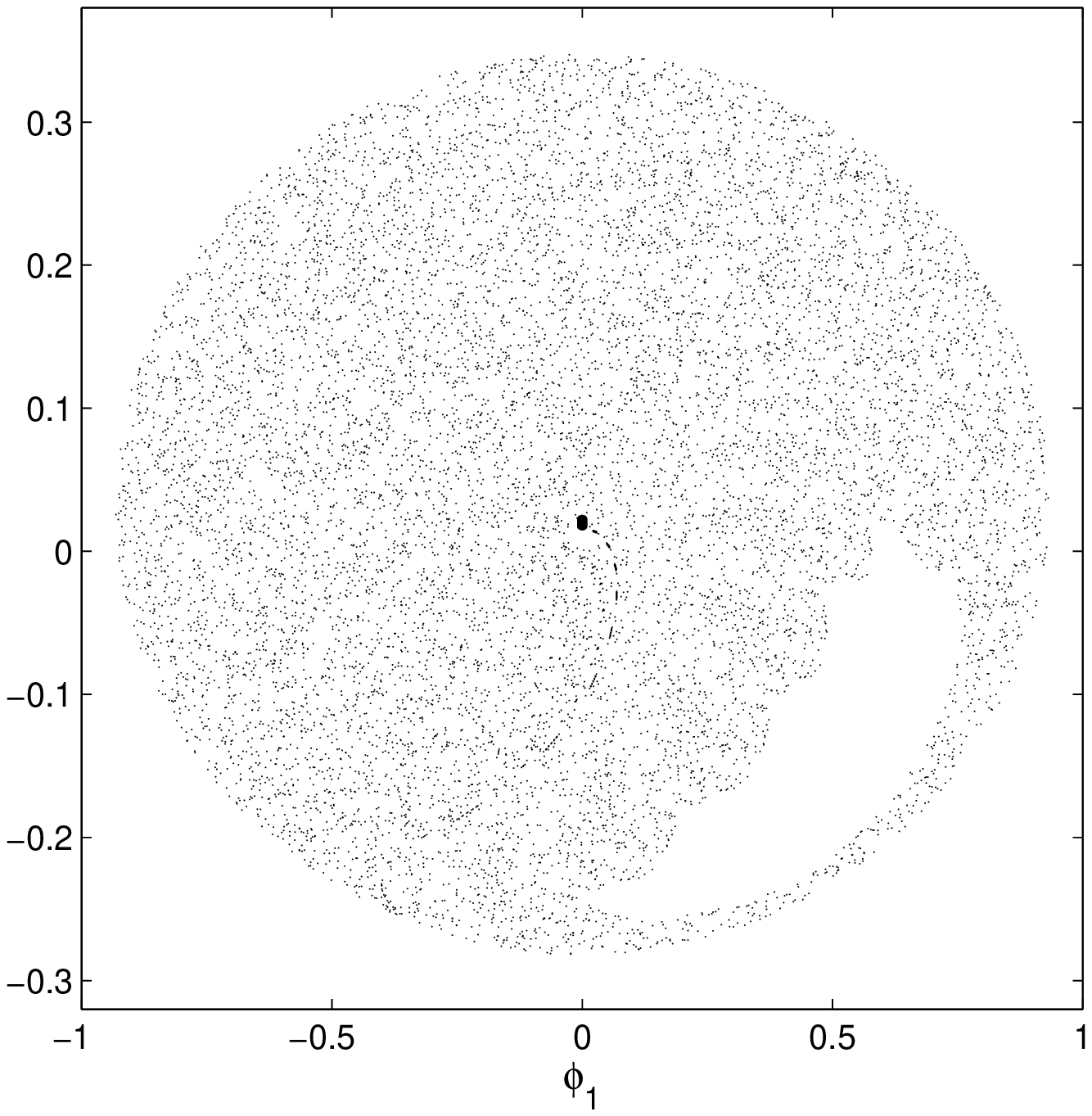,width=0.3\textwidth}&
\epsfig{figure=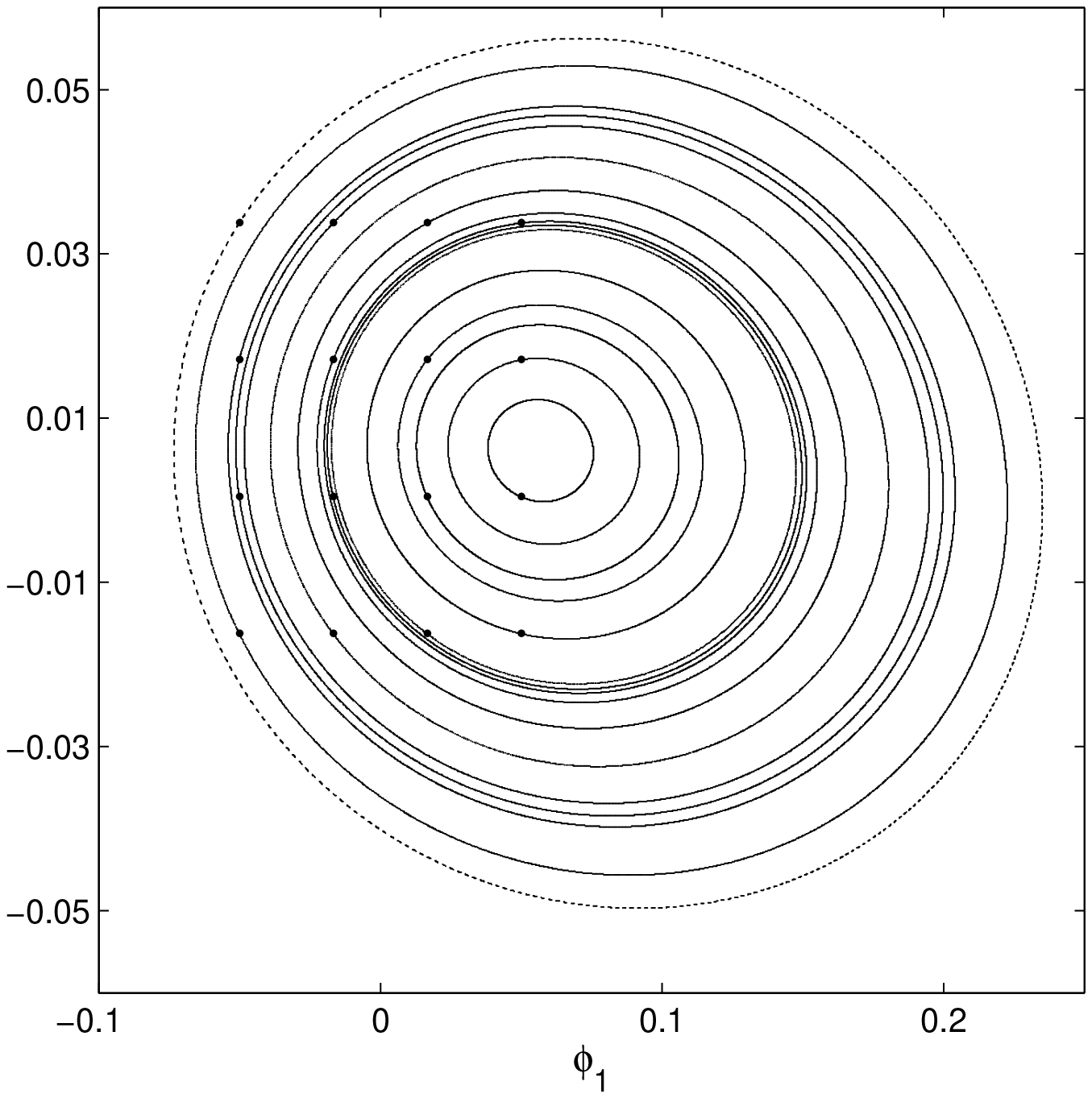,width=0.3\textwidth}&
\end{tabular}
\caption{
Poincar\'e sections of trajectories based in the vicinity of the
configurations A, B, C selected in figure~\ref{fig:Gexp}. In every case one
of the lateral condensates is almost empty, whereas the other two share
almost equally the remaining boson population, which adds up to more than
97\% of the total. The phases of the macroscopically filled condensates
are parallel to each other and opposite to that of the almost empty
condensate [such configurations are close to the reduced-phase-space
point $(\phi_1, \xi_1,\phi_2, \xi_2)=(0,0.016,\pi,-0.984)$]. The three
situations are distinguished by (from left to right)
$\tau=0.22$, $\tau=0.29$, $\tau=0.36$; in every instance
$\nu=0$. The features of the  Poincar\'e sections clearly mirror
the stability character of the relevant fixed points. Notice
furthermore that the central Poincar\'e section, relevant to the
choice falling in the unstable ``isthmus'' of figure~\ref{fig:Gexp},
covers a region roughly an order of magnitude wider than the lateral
sections, relevant to stable regions. }
\label{fig:Nexp}
\end{figure*}

An instability outbreak similar to that discussed above is found in the
neighborhood of the ``instability isthmus'' of the non-dimeric stability
diagram~\ref{fig:NS}. Indeed, the possibly present upper branch of the
non-dimeric $\tau$ curve always crosses such unstable region. This is
the case of four of the five curves plotted in figure~\ref{fig:NS} and
of the curve with $\nu=0$ plotted in figure~\ref{fig:Gexp}. The inset of
figure~\ref{fig:Gexp} is a magnification showing fixed points $A$, $B$
and $C$, located in the proximity of the instability ``isthmus'' for three
slightly different values of $\tau$  ($\tau_{\rm A}=0.22$, $\tau_{\rm B}=0.29$
and $\tau_{\rm C}=0.36$). These fixed points correspond to (slightly) different values of coordinate $\theta$, thus the corresponding configurations are quite similar to each other. In these three cases more than $97\%$ of the total
population is almost equally shared by the central and one of the lateral
condensates. The phases of these condensates are parallel to each other 
and opposite to that of the remaining, almost empty, condensate. This kind
of configuration  corresponds to the neighborhood of the reduced-phase-space
point with $(\phi_1, \xi_1,\phi_2, \xi_2) = (0,0.016,\pi,-0.984)$.
Figure~\ref{fig:Nexp} shows the Poincar\'e sections based on a set of
points closely surrounding it, for the three choices of $\tau$ listed
above, $\tau_{\rm A}$, $\tau_{\rm B}$ and $\tau_{\rm C}$.
%
%
The first and last choice --corresponding to trajectories based in the
neighborhood of a stable fixed point-- are both characterized by extremely
regular sections where the populations ``flicker'' about their initial
configuration.
In the remaining case, the fixed point belongs to the 
``unstable isthmus'' (see figure~\ref{fig:Gexp}), and the consequent 
irregular trajectories produce a quite different pattern. 
Also in the present case, a slight change in the parameters 
destroys the steady, flickering populations.
Once again the Poincar\'e sections are indistinguishable 
from one another and densely fill an area considerably 
larger than in the previous cases.

\subsection{Instability suppression due to fixed-point disappearance}
\label{S:Instsupp}
\begin{figure*}
\begin{tabular}{ccc}
\epsfig{figure=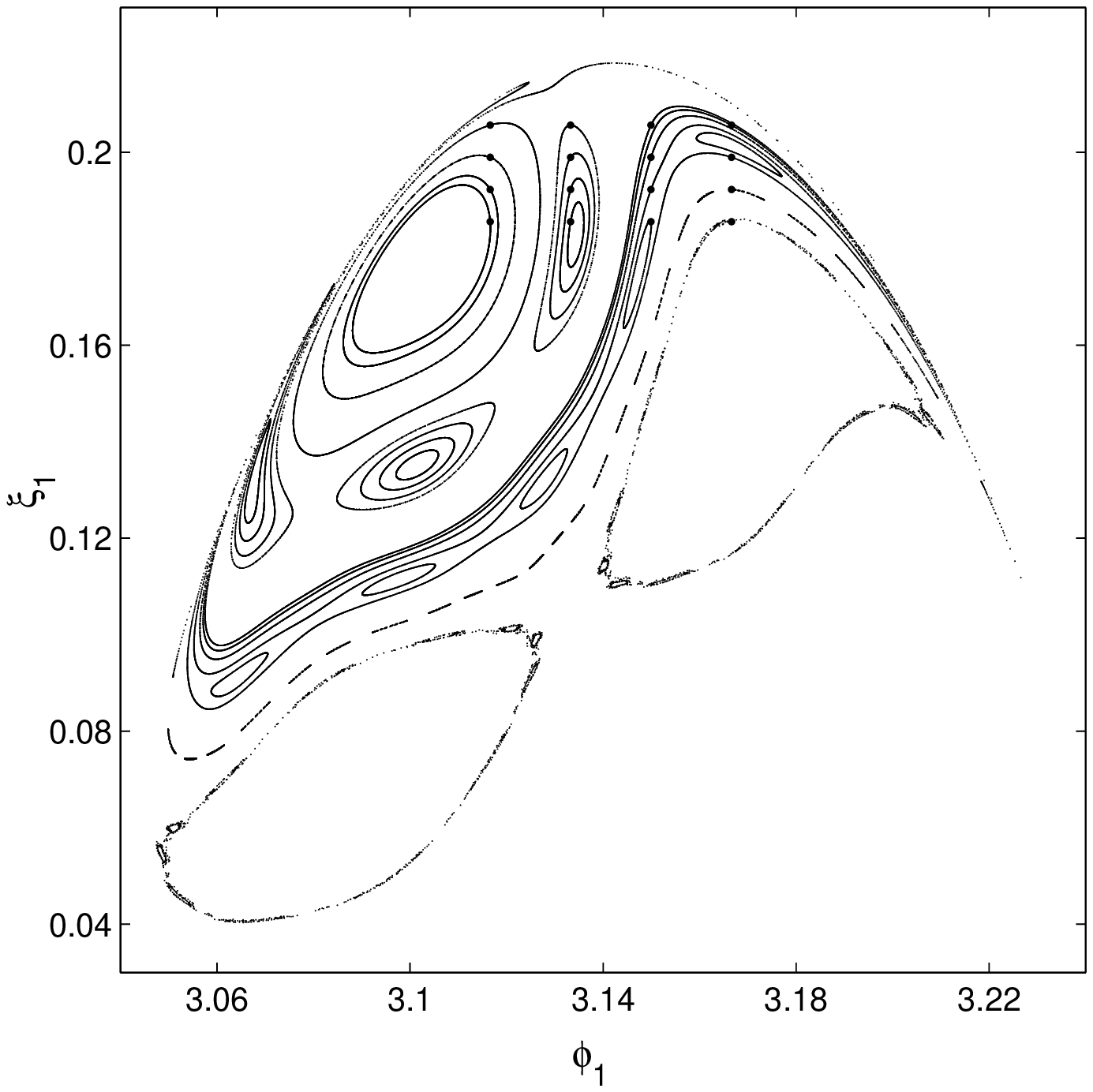,width=0.3\textwidth}
\epsfig{figure=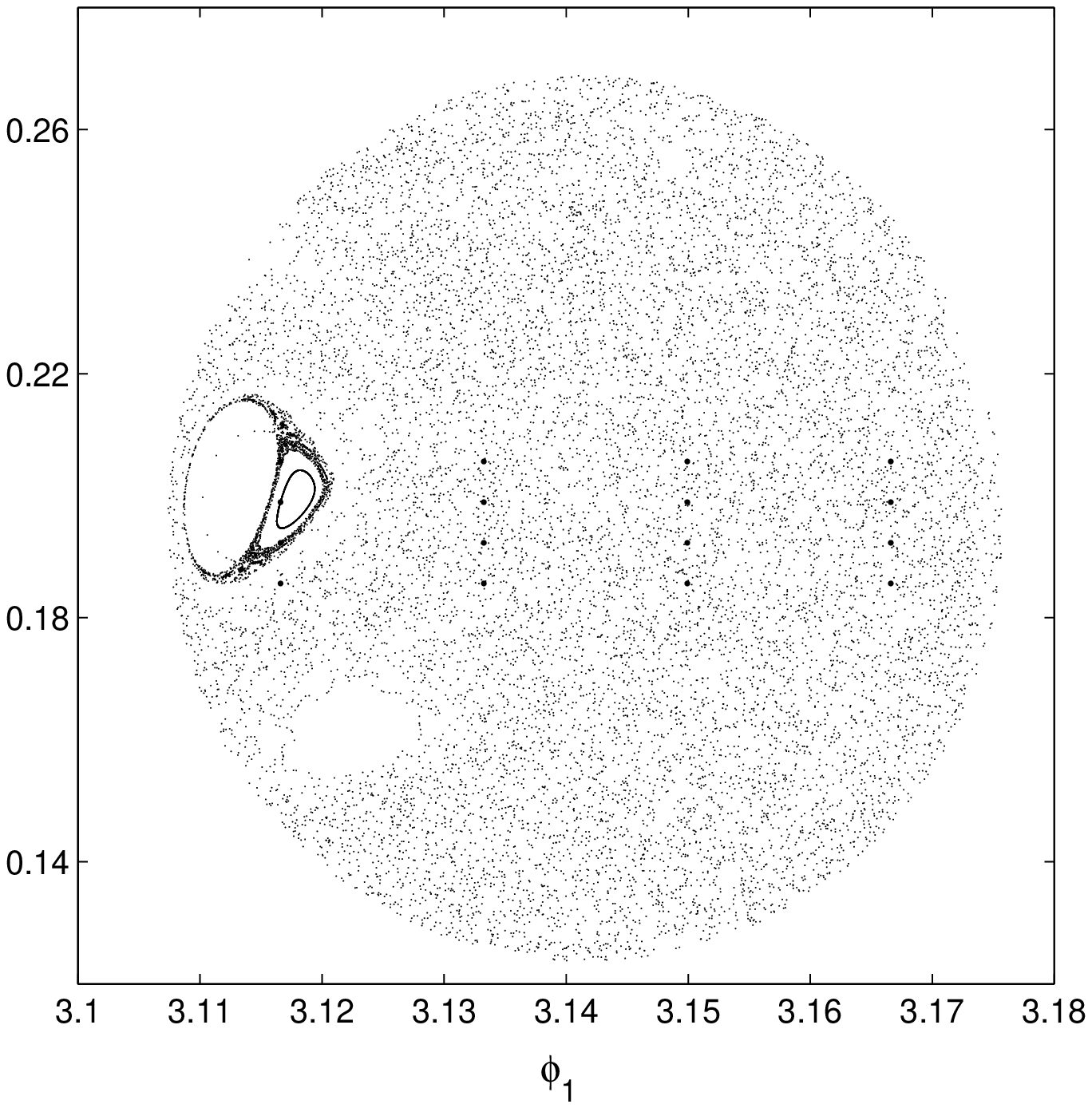,width=0.3\textwidth}&
\epsfig{figure=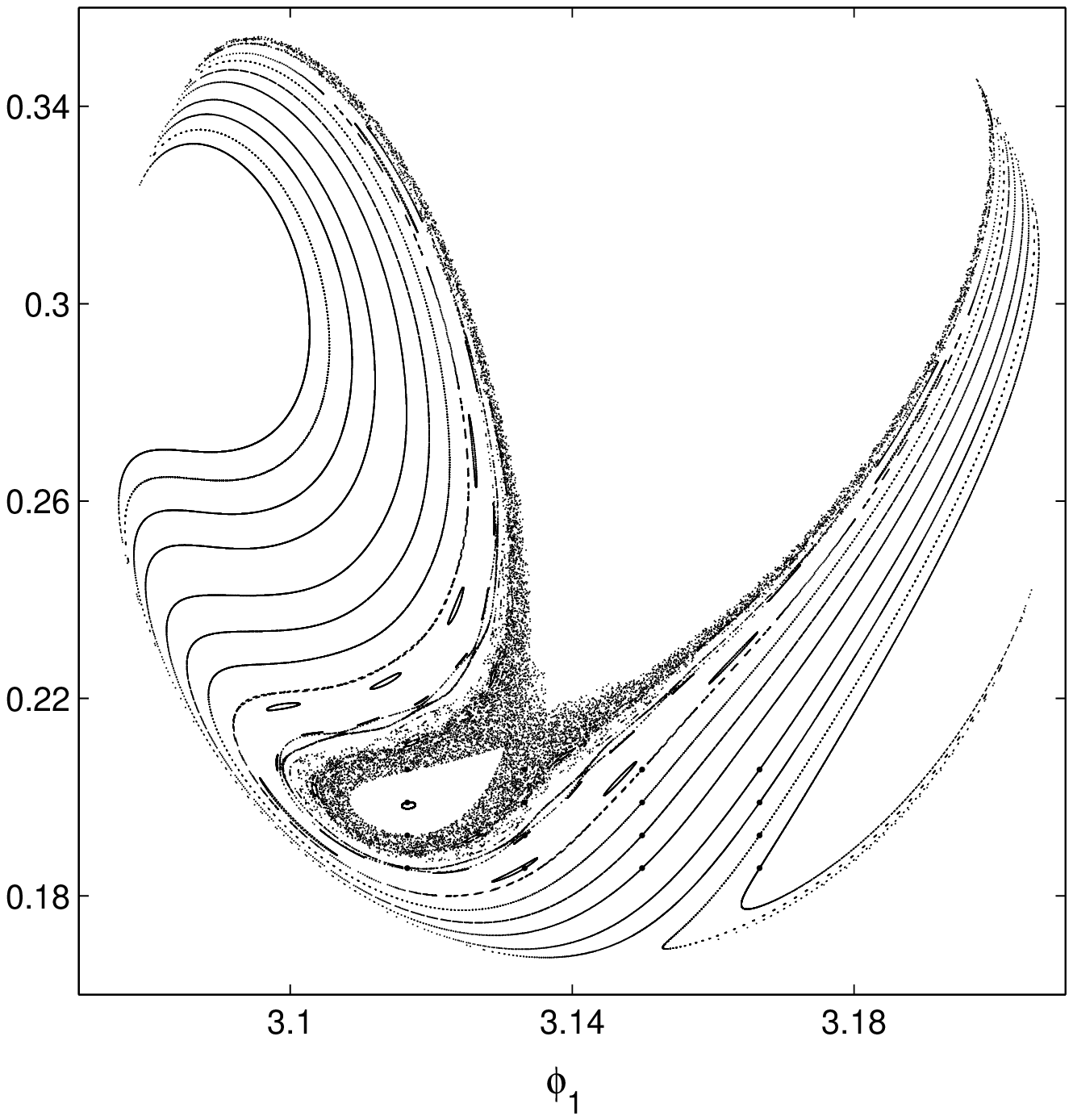,width=0.3\textwidth}&
\end{tabular}
\caption{
Regularization of the phase-space due to the disappearance of an
unstable fixed point. All of the three sets of Poincar\'e sections
refer to $\nu = 0.52$, and have been obtained from trajectories based
in the vicinity of the same configuration. For $\tau=1.66$ such
configuration is an unstable saddle, and produces quite irregular
trajectories, as shown in the central panel. Slight variations of
the hopping amplitude cause the disappearance of the unstable fixed
point, and this in turn results in an increase in the regularity of
the trajectories based in the vicinity of the corresponding configuration
(see discussion in section~\ref{S:Instsupp}). This is clearly visible in the
leftmost and rightmost panels of the figure, pertaining to $\tau=1.5$ and
$\tau=1.82$, respectively.}
\label{fig:bif}
\end{figure*}

So far, we have examined two situations where the change in the 
phase-space structure is associated with a modification in 
the stability character of a fixed point. Nevertheless,
such a change may be also related to the disappearance of a 
fixed point.
When $\nu=0.52$, e. g., a consistent portion of the lower branch of 
the non-dimeric $\tau$-curve falls in the forbidden zone,  
similar to the curve featuring the narrowest dashing style
in figure~\ref{fig:NS}. 
However, for $\tau=1.66$ (i.e. very close to the branch tip), 
both of the solutions associated to the lower branch are acceptable. 
According to the stability diagram, one of them is a maximum, whereas 
the other is an unstable saddle.
Actually, the trajectories in the neighborhood of this configuration,
produce the expected irregular Poincar\'e sections, displayed 
in the central panel of figure~\ref{fig:bif}. As $\tau$ is decreased, 
the unstable solution moves toward the forbidden zone 
(see figure~\ref{fig:NS}). For $\tau=1.5$, the maximum is the 
only non-dimeric fixed point surviving. The change in the
phase-space structure associated to the disappearance of the 
unstable fixed point is quite evident in the leftmost panel of 
figure~\ref{fig:bif} that shows the Poincar\'e sections obtained from 
the same set of base points considered in the case $\tau=1.66$. 
Despite some features signalling a certain degree of chaoticity, 
the sections are much more regular and can be easily distinguished 
from each other. 
The unstable fixed point, as well as the stable one, 
disappears also if $\tau$ is increased by a sufficient amount
(the disappearance of fixed points takes place at the branch tip 
after a coalescence process).
Similar to the previously examined 
situation, an increase in the regularity of the Poincar\'e sections 
is observed for $\tau=1.82$, when only dimeric fixed points are present
(rightmost panel of figure~\ref{fig:bif}).

\section{Phase-triggered population oscillations near 
the CDW configuration}
\label{S:CDWI}

The last case considers several unexpected and interesting 
features of CDW fixed points. Remarkably, these fixed points
are always present independently from the value of the parameters
$\tau$ and $\nu$, and have a simple configuration where the entire
boson population is equally shared between the lateral condensates,
being the phases opposite to each other.
Figure~\ref{fig:CDW} shows that the CDW are saddle points and are
expected to exhibit a stable character in extended regions of the
parameter plane. This fact, is widely confirmed by our numeric
simulations where, for initial configurations sufficiently
close to the CDW fixed point, we observe a regular dynamics 
exhibiting (possibly quite complex) periodic oscillations of 
the three condensate populations. In particular, the lateral
condensates cyclically exchange small amounts of bosons, so that their 
population imbalance oscillates. Whereas the central condensate,
whose population oscillates with a considerably smaller amplitude 
and a frequency twice than that of the lateral condensates,
participates in the dynamics acting as a bridge between the lateral ones.

\begin{figure}[htpb]
\centering 
\epsfig{figure=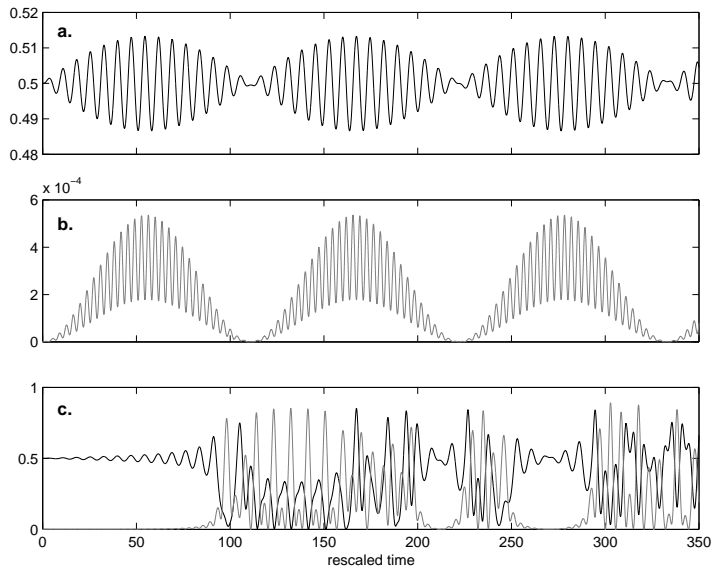,width=0.6\textwidth}
\caption{
Evolution of an initial configuration slightly displaced from the CDW, 
${\bf z}=(1/2,0,1/2 e^{i (\pi-\varphi)})$,  $\varphi=5\cdot 10^{-3}$.
Black and grey solid lines represent $n_1/N$ and $n_2/N$, respectively.
Panels {\bf a} and {\bf b} refer to the parameter choice $\nu=0$,
$\tau = 0.703$, belonging to a stable region but very close to the
border dividing it from an unstable region (see 
figure~\ref{fig:CDW} and (\ref{E:cdwB})). 
The evolution displayed in panel {\bf c} was obtained increasing 
$\tau$ by 0.005. This was sufficient to cross the border and enter 
the unstable region. The predicted instability of the CDW is 
confirmed by the strongly irregular character of the dynamics, 
which fully shows up after an apparently regular transient. 
The remaining population, $n_3$, (not shown) can be found 
exploiting the conservation of the total number of bosons 
$\sum_j n_j=N$. In the regular case, panels {\bf a} and 
{\bf b}, it essentially balances the oscillations of 
$n_1$, $n_2$ being almost vanishing.}
\label{fig:cdwsu}
\end{figure}

\begin{figure}[htpb]
\centering 
\epsfig{figure=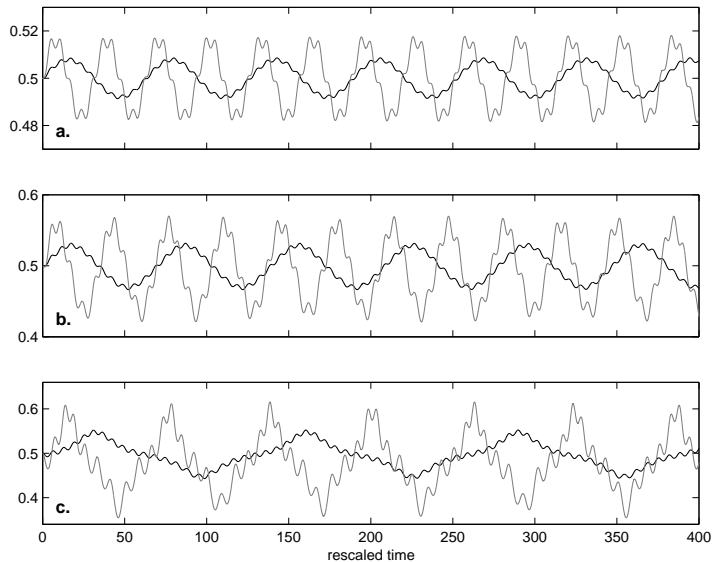,width=0.6\textwidth}
\caption{
Oscillation of the population $n_1$ relevant to initial configurations
where the total number of bosons is equally shared by the lateral condensates.
The three panels  refer to different (increasing) values of the phase 
difference  $\theta_{31}=|\theta_3-\theta_1|$. {\bf a.}  
$\theta_{31}= 9/10 \pi$;  {\bf b.}  $\theta_{31}= 1/2 \pi$; {\bf c.}  
$\theta_{31}= 1/10 \pi$. Black and grey lines refer to $\tau=0.1$ 
and $\tau = 0.2$, respectively. In every case there was no bias on 
the central condensate ($\nu=0$). Notice that the oscillation 
amplitude increases with increasing  $\theta_{31}$. Notice 
also the incipient anharmonic shape of the oscillations for 
small $\theta_{31}$ (panel {\bf c.})  }
\label{fig:cdwpops}
\end{figure}

Figure \ref{fig:cdwsu} shows the 
evolution of a configuration quite close to the CDW for two 
parameter choices belonging to different stability regions 
of figure~\ref{fig:CDW}.
%
%
As expected, in the stable case (panels \textbf{a} and \textbf{b}),
the amplitude of the population 
oscillations increases with increasing displacement of the initial 
configuration from the CDW. Of course, in general, such displacement 
cannot be increased indefinitely: the initial configuration would 
eventually leave the ``domain of influence'' of the CDW and the 
consequent trajectory would be determined by the stability 
character of a different fixed point. 
The same considerations apply to configurations obtained 
from the CDW by setting the phase difference between the lateral
condensates to a value $\theta_{31}=|\theta_3-\theta_1|\ne \pi$ 
and maintaining the total boson population equally shared
by the lateral condensates.
In this case the oscillation amplitude is a function of
$|\theta_3-\theta_1|$ with a minimum at $|\theta_{31}|=\pi$, where
it vanishes. Our numeric simulations confirm these predictions.
In fact, for sufficiently small $\tau$'s, regular oscillatory 
dynamics is observed over the entire range of $\theta_{31}$. This 
is clearly shown in figure~\ref{fig:cdwpops}, displaying the population 
oscillations pertaining to initial configurations identified by 
several values of $\theta_{31}$. 
%

\begin{figure}[htpb]
\centering 
\epsfig{figure=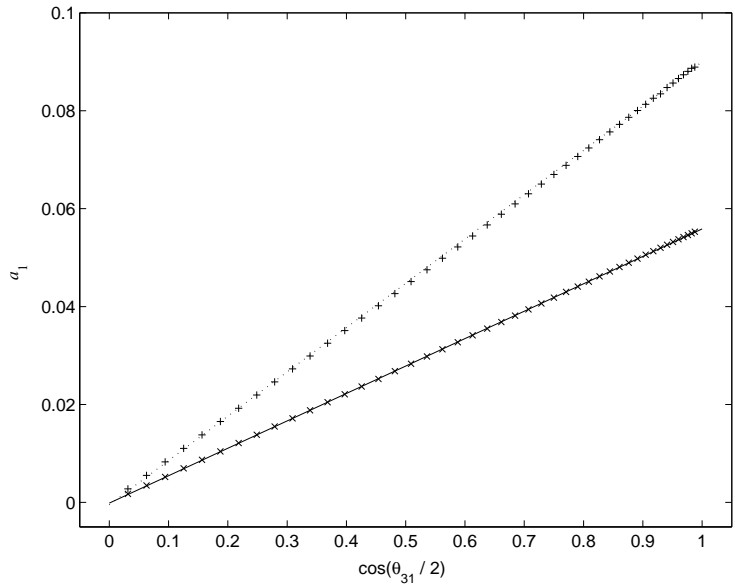,width=0.5\textwidth,angle=270
}
\caption{
Amplitude of the population oscillations 
of one of the lateral condensates for different values of the 
initial phase difference $\theta_{31}$ and $\nu=0$.
Crosses ($\times$): $\tau=0.1$; Plus signs ($+$): $\tau=0.15$. The straight 
lines are linear fits of the two sets of data.}
\label{fig:xpamp}
\end{figure}

\begin{figure}[htpb]
\centering 
\epsfig{figure=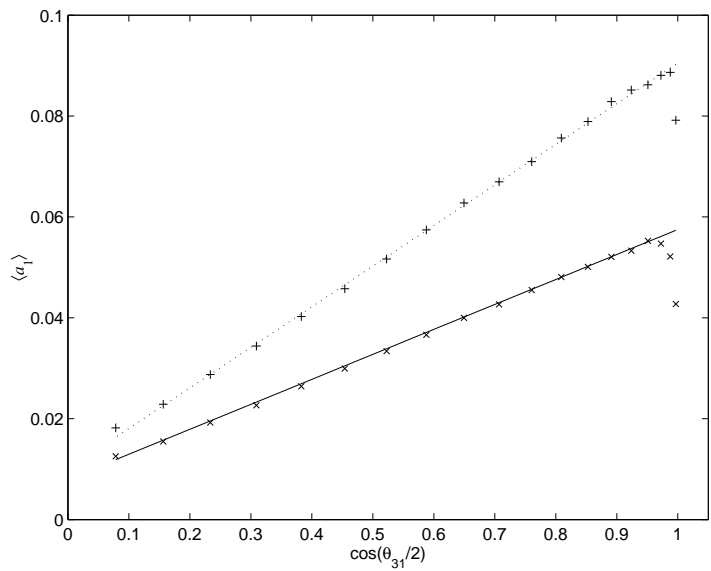,width=0.6\textwidth}
\caption{
A simple dependence of the population oscillation on the initial phase
difference $\theta_{31}$ is present even if the initial population
configuration does not reproduce that of the CDW exactly. Each of the
plotted data was obtained  averaging the results of 300 simulations where
the initial populations were allowed to differ by about 1\% from the
configuration $(1/2,0,1/2)$, and the phase of the central 
(almost depleted) condensate was chosen randomly.
}
\label{fig:xpav}
\end{figure}

%
%
A further important feature of the CDW fixed points is highlighted by
figure~\ref{fig:xpamp} showing that for small values of $\tau$,
the amplitude $a_1$ of the oscillation 
of the population $n_1$ exhibits a quite simple dependence on 
the phase difference $\theta_{31}$ of the relevant initial 
configuration. More precisely $a_1$  linearly depends on 
$\cos (\theta_{31}/2)$. A rigorous discussion of such 
dependence is beyond the scope of the present article. 
However, some insight in this sense can be gained by
adopting an approach similar to that developed in ~\cite{Lphys3}
in the study of the stability of the dimeric integrable subregime.
%
%
%
Interestingly, the same linear behavior as in figure~\ref{fig:xpamp} is 
obtained by relaxing the requirement that the entire boson population 
is equally shared by the two lateral condensates. This is clearly 
shown in figure~\ref{fig:xpav}, where we plotted the average oscillation 
amplitude relevant to initial configurations characterized by a 
certain phase difference $\theta_{31}$ and condensate populations 
$n_j$ slightly displaced from that of the CDW fixed point. 

The plots displayed in figures~\ref{fig:xpamp} and \ref{fig:xpav} 
suggest the possibility of measuring the phase difference 
between two equally populated condensates.
Let us stressing the fact that, in order to profitably exploit this
quantum effect to measure the difference of phase between
two condensates, we need of a nondestructive real-time probing of the 
condensates such as the technique 
employed to image dimer-like condensates confined by magnetic traps 
\cite{Andrews97,Andrews}.

\section{Conclusions}
\label{S:conc}

The present paper has been inspired by the wish of gaining a deeper
insight both on the complexity of three-well dynamics (which, as noted
above, is the paradigm of the behavior one can expect in more complex
BECs arrays) and on the possibility of detecting macroscopic effects
amenable to experimental observation.

Recently, it has been proposed that the superposition of multiple
laser beams could allow to create unusual optical lattices in a
controlled manner \cite{superL}.
In this respect, we observe that a suitable choice of laser
wavelengths could generate a superlattice whose elementary cell is
characterized by a three-well potential such as the one addressed in
the present paper. The further minimization of the tunneling between
subsequent trimeric cells should provide an alternative experimental
realization of the system under concern. A nice example in this sense is
the trimerized Kagom\`e optical lattice~\cite{Santos}
providing (multiple copies of) the
trapping potential typical of a symmetric closed trimer.

In the present paper we focus on the rich phenomenology offered by the
trimer and on the possibility of triggering macroscopic effects
through a suitable choice of the adjustable parameters, relegating as
much as possible to the appendices the technical details necessarily
involved in a systematic analysis. 
The macroscopic effects we evidence in this paper are obtained through
a careful analysis of the stability diagrams introduced in \cite{PFLet1}, 
whose operational value is reviewed in section \ref{S:SD}.
Their derivation is thoroughly discussed in \ref{S:ASD}. 
Such diagrams describe the trimer
phase space in terms of the stability character of
dynamical fixed points for any choice of the significant
parameters. This wealth of information enriches the recent
literature about the dynamical stability of mesoscopic BEC arrays
\cite{Wu}-\cite{Kalosakas02}, \cite{STrimer}-\cite{Chong}, \cite{Menotti03} 
%
and supplies an essential tool in designing
experimental realizations of the trimer dynamics. Also,
at the theoretical level, it provides a sound basis for gaining a deeper
insight into the quantum counterpart of classical dynamical
instability, which is far from being fully understood.

The use of stability diagrams leads to identify in section
\ref{S:macroeff} three different mechanisms able to modify in a
substantial way the trimer dynamics. 
These are exploited in sections \ref{S:priming}, \ref{S:losscoher} and
\ref{S:Instsupp} to prime instability in breather-like regular oscillations,
loss of coherence in anti-breather states, and suppression of instability
due to fixed-point disappearance, respectively. Macroscopic dynamical
oscillations are shown to be triggered by means of slight parameter
changes thus providing a stimulating basis for planning actual experiments
in the scheme described in~\cite{Anker05}-\cite{Oberthaler_mt}.
The analysis of the stability diagrams in section \ref{S:CDWI}
also evidences that a simple configuration exists 
allowing to relate in a direct manner the
initial phase difference between two condensates to the amplitude of
the ensuing population oscillations.  This suggests a rather simple
scheme aimed at testing experimentally the limits of validity of the
mean-field picture, where each condensate is endowed with a
macroscopic phase.
The study of the fixed points --inherent in the classical picture--
reported in sections \ref{S:Dyn} and \ref{S:SD} proves useful also in
unraveling the complex structure of the energy spectrum of the quantum
trimer, in that it allows to recognize a (either exact or approximate)
mapping with a simpler dimeric system \cite{MFPersist}.  The mapping
is shown to persist also at the quantum level, thus extending the
results found in \cite{Aubry} to the case of the quantum
trimer.  These results provide a useful starting point in
investigating the quantum counterpart~\cite{Kol} of classical dynamical
instability.  A deeper insight into the relation between the classical
and quantum picture is certainly useful for the approaches to this
issue based on the dynamical algebra method \cite{FP1,Zhang,Lphys3}
or on the measure of quantum-state entanglement \cite{entang}. 
%
%
%

\begin{appendix}

\section{Derivation of the stability diagrams}
\label{S:ASD}

The linear stability character of a fixed point
${\mathbf x}^t=(x_1,x_2,x_3)$ can be determined 
by studying the eigenvalues of the matrix associated to the (linearized) 
equations governing the dynamics of the (small) displacements 
${\mathbf v}^t= ({\mathbf q}^t,{\mathbf p}^t)$
from the fixed point itself, where 
${\mathbf z}^t= (z_1, z_2, z_3) = 
\sqrt N ({\mathbf x} + {\mathbf q}^t+i\, {\mathbf p}^t )$. The Hamiltonian, up to
the second order in the small displacements ${\mathbf v}^t$, reads
\begin{equation}
\label{E:Hpq}
{\cal H}({\mathbf z})-\chi N = {\cal H}(\sqrt N {\mathbf x})+
U N^2 \frac{\tau}{2} {\mathbf v}^t  M_h  {\mathbf v},
\end{equation}
where 
\begin{equation}
\label{E:mqp}
M_h=\left(
\begin{array}{cc}
Q & 0 \\
0 & P
\end{array} \right)
, \,\,\,
Q= \left | P_{h k}+\frac{8 x_j^2}{\tau}\delta_{h k} \right | 
, \,\,\,
P=\left(
\begin{array}{ccc}
\Delta_1 & -1 & 0\\
-1 & \Delta_2 & -1\\
0 & -1 &\Delta_3
\end{array} \right)
\end{equation}
and $\Delta_j = 2(2 x_j^2 -m -\delta_{j\,2}\, \nu)/\tau$.
As it is well known, linear instability occurs when 
at least one of such (in principle complex) eigenvalues features a 
positive real part.
The linearized equations obtained from Hamiltonian (\ref{E:Hpq}) 
through Poisson's brackets
$\{q_j,p_k\}=\,\frac{\delta_{j k}}{2\,N\,\hbar}$
(these stem from the original ones $\{z_j,z^*_k\}=\frac{i\delta_{j k}}{\hbar}$)
have the form
\begin{equation}
\label{E:M11}
\dot {\mathbf v} =\{{\mathbf v},{\cal H}\} =
\frac{U\, N\,\tau}{2\,\hbar} M_s {\mathbf v},\qquad 
M_s=\left(
\begin{array}{cc}
0 & Q \\
-P & 0
\end{array}
\right).
\end{equation}  
In accordance with the expected time reversal symmetry of the linear 
dynamics, the six (complex) eigenvalues of matrix $M_s$ come in pairs 
featuring the same modulus and opposite signs. More precisely each 
pair of eigenvalues of $M_s$ consists in the signed square roots 
of an eigenvalue of the three-by-three matrix $-Q P$. Indeed, due to 
the diagonal block structure of $M_s$, the relevant eigenvalue equation is
%
${\cal D}(\lambda) 
=\det(\lambda I-M_s)=\det(\lambda^2 I+QP) = \lambda^2(\lambda^4-s \lambda^2+p)=0$,
where in writing the characteristic polynomial of $-Q P$ we took into 
account the fact that, similar to $P$, such a matrix has a zero eigenvalue.
This feature is the consequence of the fact that both
${\mathbf q}^t$ and $ {\mathbf p}^t$ are constrained by the condition
$|z_1|^2 + |z_2|^2  +|z_3|^2= N $.
According to ${\cal D}(\lambda)=0$ the fixed point is (linearly) stable 
provided that the three conditions
\begin{equation}
\label{E:st_cond}
s^2-4 p \geq 0 ,\qquad s \leq 0,\qquad p \geq 0
\end{equation}
are simultaneously met. Indeed in this case the two non-zero 
eigenvalues of $-Q P$ are real and non-positive, so that  the 
four non-zero eigenvalues of $M_s$  --- namely their signed 
square roots --- are purely imaginary. 
Conversely, whenever an eigenvalue of $-QP$ is complex or 
real positive, one of its square roots has necessarily a 
positive real part, giving rise to (linear) instability.
The following subsections describe in some detail the 
construction of the stability diagrams for the DFP, 
the NFP and the CDW.

\subsection{Dimeric fixed points}
\label{S:DR}
As illustrated in section \ref{S:Dyn}, the dimeric fixed points (DFP)
pertain to the integrable dimeric subregime \cite{Lphys3} of 
the dynamics described by system (\ref{E:dyneq}), which is 
characterized by $z_1(t)\equiv z_3(t)$.  
In this situation the fixed point configuration $(x_1, x_2, x_3)$
is described by 
%
$x_1(\theta) =x_3 (\theta)= \cos \theta/{\sqrt 2}$ 
and $x_2(\theta)= \sin \theta$ (see Eqs. (\ref{dfp})), 
where the tangent
${\alpha}(\theta) =\tan(\theta) ={x_2}/({\sqrt 2 \, x_1})$
must be a real root of polynomial
%
${\cal P}({\alpha};\tau,\nu)\equiv {\alpha}^4+
A_{\nu} \tau^{-1} {\alpha}^3- B_{\nu} \tau^{-1} {\alpha}-1 
$
(see formula (\ref{dfp_c})) with 
$A_{\nu} =\sqrt 2(2-\nu)$, $B_{\nu} =\sqrt 2(1+\nu)$.
Before proceeding to a more detailed discussion of the DFP case, 
we remark that the previous polar 
representation of $x_j$'s allows one to readily
relate variable $\theta$ to the fixed point $(x_1, x_2, x_3)$.
Particularly, $\tan(\theta) ={x_2}/({\sqrt 2 \, x_1})$
shows how $\theta \approx 0$ corresponds to a situation where the 
central condensate is almost depleted, so that the total 
population is (equally) shared by the lateral condensates. 
Conversely, when $\theta\approx \pm {\frac{\pi}{2}}_\mp$ the 
lateral condensates are almost empty and the total population 
is concentrated in the central one. Notice also that the sign 
of $\theta$ is similarly strictly related to the phase 
differences between the central condensate and the lateral 
ones. More precisely such phase differences are 0 or $\pm \pi$ 
depending on whether $\theta$ is positive or negative.

As we discussed above, solving the fixed point equations 
for the DFP amounts to finding the real roots of the 
fourth-degree polynomial ${\cal P}({\alpha};\tau,\nu)$ 
(henceforth denoted by $P_D({\alpha};\tau;\nu)$). Four real solutions 
are found from $P_D({\alpha};\tau;\nu)=0$ for
\begin{equation}
\label{E:Dlobe}
\left|\nu-{1}/{2}\right|<\, {1}/{2}, \quad
\tau\leq\frac{1}{\sqrt 2} \left[\left({3}/{2}\right)^{{2}/{3}}
-\left|\nu-{1}/{2}\right|^{{2}/{3}} \right]^{{3}/{2}} 
\end{equation}
and two real solutions elsewhere in the parameter plane, 
as it is illustrated in figure \ref{fig:DN}.
In section \ref{S:SD}, we showed how
equation $P_D({\alpha};\tau;\nu)=0$ suggests that, 
for any choice of parameters $\tau = \tilde \tau$, $\nu= \tilde \nu$, 
the DFP can be found in the $\theta-\tau$ plane from the 
intersection of the parametric curve (see \ref{E:GtauC})
\begin{equation}
\label{E:Dpc}
\tau = \tau_D({\alpha};\nu) = \sqrt 2\, {\alpha}\, 
\frac{(\nu-2){\alpha}^2+(\nu+1)}{{\alpha}^4-1}\, ,\quad \nu= \tilde \nu\, ,
\end{equation}
with the vertical straight line $\tau=\tilde \tau$.
The curve $\tau_D(\alpha;\tilde \nu)$ features two asymptotes 
at $\theta=\pm \pi/4$ (${\alpha}=\pm 1$). This is in agreement 
with the fact that polynomial $P_D({\alpha};\tau,\nu)$ has no 
less than two real solutions. In particular it is easy to verify 
that when $\nu=1/2$ the values $\theta=\pm 1/2$ are solutions 
of $P_D({\alpha};\tau;\nu)=0$. Hence, in this case, the two unbounded 
branches  coalesce with their asymptotes, thus providing 
$\tau$-independent solutions corresponding to the 
configurations $(x_1,x_2,x_3)=(1/2,\pm 1/\sqrt 2 ,1/2)$.
More generally, when $\nu\neq 1/2$ the branches of the curve 
$\tau_D(\alpha;\tilde \nu)$, and hence the roots of 
$P_D({\alpha};\tau;\nu)=0$, are characterized by a reflection symmetry 
with respect to the nearest asymptote (see figure~\ref{fig:DS}). 
This can be seen by  noticing that
\begin{equation}
\tau_D \left [ \tan \left( \vartheta \pm \frac{\pi}{4}
\right );\bar\nu +\frac{1}{2}
\right ] 
= \frac{2\,\bar\nu\mp3 \sin(2\vartheta)}{\sin(2\vartheta)} 
\frac{\cos(2\vartheta)}{2\sqrt 2}\, , \, 
\vartheta \in \left [-\frac{\pi}{4},\frac{\pi}{4} \right ]
\end{equation}
where the auxiliary angle $\vartheta$ describes 
the positive (negative) solutions $\theta = \vartheta+\pi/4$  
($\theta = \vartheta-\pi/4$). One easily verifies that
$\tau_D \bigl [ \tan(\vartheta \pm {\pi}/{4}); \bar\nu + {1}/{2}\bigr ] 
=\tau_D \bigl [\tan(\pm {\pi}/{4}-\vartheta); {1}/{2}-\bar\nu \bigr ]$.

{\it Local character of the DFP}. This information is obtained
by exploiting the constraint 
$x_j \neq 0$ and the fixed point equations (\ref{E:fpsys}). After
expressing the quantities appearing in matrices $Q$ and $P$ 
(see (\ref{E:mqp})) as
$\Delta_1=\Delta_3={x_2}/{x_1}=\sqrt 2\, {\alpha}$
and $\Delta_2=(x_1+x_3)/{x_2}={\sqrt 2}/{\alpha}$,
the analysis of the signature of both matrix $Q$ and $P$ depending
on parameter $\alpha= \tan \theta$
leads to the following conclusions.
For any choice of the parameters $\tau$ and $\nu$, 
a DFP is a minimum if the corresponding angular coordinate, 
$\theta$, is positive. If, conversely, $\theta<0$ the fixed 
point is either a maximum or a saddle point depending on 
whether $\tau$ is greater or less than $d(\tan \theta)$
where $d({\alpha})=\max\left[d_1({\alpha}),d_2({\alpha})\right]$
with $d_1({\alpha})=-2{\sqrt 2}/[ \alpha (1+{\alpha}^2) ]$
and $d_2({\alpha})=$ $-6{\sqrt 2}/[ \alpha^3 (1+{\alpha}^2)^3 ]$.
Combining such results with the linear stability analysis based on
equation ${\cal D}(\lambda)=0$ (defined after formula (\ref{E:M11}))
confirms that both maxima and minima are, as expected, stable fixed
points, whereas all the saddles of the dimeric regime are unstable.
The situation for the DFP is summarized in the corresponding 
{\it stability diagram} displayed in figure~\ref{fig:DS}. The 
differently shaded region pertain to fixed points with a different 
stability character. 

\subsection{Non dimeric regime}
\label{S:NR}
Since in this case $x_1\neq x_3$, the polar representation of 
nondimeric fixed points (NFP) is not as simple as that of the
DFP case. Still, one can take advantage from the conservation
of the total boson number. Some algebraic manipulations of the
fixed-point equations (see {\it non-dimeric solutions} in
section \ref{S:FP}) suggest to switch to the new coordinates 
$X_1=$ $(x_1+x_3)/\sqrt 2$, $X_2=x_2$ and
$X_3=(x_1-x_3)/{\sqrt 2}$,  
obeying the constraint 
$X^2_1+X^2_2+X^2_3=1$, before setting
$X_1 = R \cos \theta$ and  $ X_2= R  \sin \theta$.
This way $X_3= \pm\sqrt{1-X_1^2-x_2^2}$ and radial coordinate 
(\ref{ndfp_ra}) are shown to depend on $\theta$ and $\tau$ as 
\begin{equation}
\label{E:RG}
R({\alpha};\,\tau)=
\sqrt{
\frac{(1+\alpha^2)(2\sqrt 2-\alpha\, 
\tau)}{2\sqrt 2 (2+\alpha^2)}
}
, \,
X_3(\alpha)\!=
\pm\sqrt{\frac{4\!+\!\sqrt 2 \alpha(1+\tau \alpha^2)}{4(2+{\alpha}^2)}}
\end{equation}
with ${\alpha}=\tan \theta ={X_2}/{X_1}$, and the fixed points equations
are equivalent to the equation for the roots of polynomial
${\cal P}({\alpha};\tau,\nu)\equiv {\alpha}^4+ A_{\nu} \tau^{-1} {\alpha}^3- 
B_{\nu} \tau^{-1} {\alpha}-{4}/{3} =0 $,
(henceforth denoted by $P_N({\alpha};\nu, \tau)$), where 
$A_{\nu}={2 \sqrt 2(2-\nu)}/{3}$, 
$B_{\nu}={4\sqrt 2(1+\nu)}/{3}$ (see \ref{ndfp_pol}). 
Recalling that we are ultimately looking for real solutions of system
(\ref{E:fpsys}) it is clear that the roots which make $R({\alpha})$ or
$X_3(\alpha)$ imaginary must be discarded. 
Before analysing how NFP are identified, it is important to remark that 
in the NFP case the relation among the auxiliary variable $\theta$ and
the configuration of the fixed point is not as transparent as in the DFP
case, basically due to the $\theta$ dependence of the radial variable
(\ref{E:RG}). Nevertheless, within the NFP stability diagram some special 
regions can be still recognized. Notice indeed that, according to the
previous definitions of $X_1$, $X_2$ and $X_3$ as well as to (\ref{E:RG}),
the (forbidden) limit situations $X_3\to0$ and $R\to0$ correspond to the
dimeric case $x_1=x_3$ and to the CDW case $x_1=-x_3$, $x_2=0$,
respectively.
As we shall discuss below, these limit situations are actually the boundaries
of two forbidden regions in the NFP diagram (see (\ref{E:forb}) and discussion). 
Due to the restrictions on $R$ and $X_3$, the diagram of figure~\ref{fig:NN},
giving the number of fixed points relevant to system (\ref{E:fpsys}) on varying 
parameters $\tau$ and $\nu$, is significantly richer than the corresponding
diagram for the number of real roots of polynomial $P_N({\alpha};\nu, \tau)$.
Indeed the latter consists of four regions, delimited by the solid lines
appearing in figure~\ref{fig:NN}. Two of these regions correspond to
situations where  $P_N({\alpha};\nu,\tau)$ features only two real roots,
whereas within the remaining regions the roots are either four or none. 
Despite they are real, some of these roots can make either $R({\alpha})$ 
or $X_3({\alpha})$ complex. Hence some new regions appear within the ones
delimited by the solid curves, where the number of solutions must be
decreased by either one or two units.   

As discussed in section \ref{S:SD}, the features of diagram (\ref{fig:NN})
can be better understood by resorting to the $\theta-\tau$ plane, where,
similar to the DFP case, for a given choice $\tau=\tilde \tau$, $\nu=\tilde \nu$,
the real roots of polynomial $P_N({\alpha};\nu, \tau)$ correspond to the
intersections of curve
\begin{equation}
\label{E:Gpc}
\tau_N({\alpha};\nu) =  2\, {\alpha}\, [(\nu-2){\alpha}^2+
\nu+2]/(3{\alpha}^4+1)\, , \quad \nu =\tilde\nu \, ,
\end{equation}
with the straight line $\tau=\tilde \tau$.
In the half plane $\tau>0$ such curve 
consists in general of two bounded, roughly bell-shaped, 
branches. More precisely they are defined within the $\theta$ 
intervals $[-\nu_b,0]$ and $[\nu_b, \pi/2]$, respectively, 
where $\nu_b=$ $\arctan(\sqrt{(2+\nu)/(2-\nu)})$.  When $\nu>2$ 
($\nu<-2$) the former (latter) branch, which from now on will 
be referred to as the negative (positive) branch, disappears. 
These features readily explain the four regions in the diagram 
describing the number of real roots of $P_N({\alpha};\nu, \tau)$.
Hence, provided $\tilde\tau$ is sufficiently small, there are
either four or two solutions depending on whether $|\tilde \nu|<2$
or $|\tilde \nu|>2$.
Indeed,  for any straight line at a given value of $\tau$, 
there exist for each branch a value of $\nu$ such that the 
branch and the straight line are tangent. Denoting such 
values $\bar \nu_t^\pm(\tau)$, where the superscript labels 
the branch, it is easy to check that $\mp\bar \nu_t^\pm(\tau)$ 
is always a growing function of $\tau$. Hence two solutions 
relevant to  the negative (positive) branch are always 
present for $\nu>\bar \nu_t^-(\tau)$ [$\nu<\bar \nu_t^+(\tau)$].
Notice further that the presence of a region of parameters where 
four solutions are found implies that there exists some $\tau_c>0$ 
such that $\bar \nu_t^+(\tau_c)= \bar \nu_t^-(\tau_c)$. Then
it is clear that $P_N({\alpha};\nu, \tau)$ 
features  four real roots in the region 
$\{\bar \nu_t^-<\nu<\bar \nu_t^+;\,\tau<\tau_c\}$, no real roots 
in the region $\{\bar \nu_t^+<\nu<\bar \nu_t^-;\,\tau>\tau_c\}$ 
and two real roots elsewhere. The explicit expressions of functions 
$\bar \nu_t^-(\tau)$ and  $\bar \nu_t^+(\tau)$ for 
${\alpha}<0$ and ${\alpha}>0$, respectively,
(solid lines in figure~\ref{fig:NS}) can be easily described 
in a parametric way by means of two curves 
$\tau_t({\alpha})$, $\nu_t({\alpha})$: the curve 
$\tau_t({\alpha}) =16\,{\alpha}^3/(3 {\alpha}^6+9{\alpha}^4 -3{\alpha}^2-1)$
intersects $\tau_N ({\alpha};\nu)$ at the tips of its bell-shaped branches 
while $\nu_t (\alpha)$ is such that
$\frac{d}{d{\alpha}} \tau_N ({\alpha};\nu)=0$
at ${\nu=\nu_t({\alpha})}$ 
and $\tau_N [{\alpha};\nu_t({\alpha})]=\tau_t({\alpha})$.
The requirement that $R$, $X_3 \in {\mathbb R}$ determines 
two forbidden regions in the stability 
diagram:$\{\theta<0,\;\tau>\tau_3(\tan \theta)\}$ and 
$\{\theta>0,\;\tau > \tau_R(\tan \theta)\}$, where 
\begin{equation}
\label{E:forb}
\tau_3({\alpha}) = -{2}/{{\alpha}(1+2\,{\alpha}^2)}, 
\qquad \tau_R({\alpha}) = {2}/{\alpha}\, ,
\end{equation}
(see figure~\ref{fig:NS}).
The information about the forbidden regions can be visualized
in parallel in 
the $\tau \!-\! \nu$ plane by drawing the curves 
$\bar\nu_R(\tau)$ and $\bar\nu_3(\tau)$ 
describing the intersections between $\tau_N (\alpha; \nu)$ 
and the borders of the 
forbidden regions, $\tau_R({\alpha})$ and $\tau_3({\alpha})$, 
respectively (dotted curves of figure~\ref{fig:NN}).
According to what we said so far, these curves  enclose two regions
where one of the real roots of polynomial $P_N({\alpha};\nu, \tau)$
must be excluded from the set of solutions of system (\ref{E:fpsys}):
$\nu>\bar\nu_3(\tau)$ and $\nu<\bar\nu_R(\tau)$. 
Similarly two of these real roots must be discarded in the regions  
$\{\bar \tau_R(\nu)<\tau <\tau_t^+(\nu),\,\nu<\nu_2^+\}$ and  
$\{\bar \tau_3(\nu)<\tau <\tau_t^-(\nu),\,\nu>\nu_2^-\}$. 
Notice also that $\tau_R(\nu)$ [$\tau_3(\nu)$] is expectedly 
tangent to $\bar \nu_t^+(\tau)$ [$\bar\nu_t^-(\tau)$] at 
$\tau=\tau_2^+$, [$\tau=\tau_2^+$].  

{\it Local character of the NFP}. The study of the 
signature of both matrix $Q$ and $P$ defined in (\ref{E:mqp}) 
reveals that NFPs are either maxima or saddle points depending 
on whether they lie within the region 
$\{ \theta < 0 , \,\tau_t(\tan \theta)<\tau<\tau_3(\tan \theta) \}$ 
or elsewhere in the allowed zone.
In parallel, the linear stability analysis based on 
polynomial ${\cal D}(\lambda)$ (see the discussion relevant to
(\ref{E:st_cond})) discloses the (linear) stability of most 
of the saddle points lying
in the positive $\theta$ quadrant, namely the ones belonging to the
light grey region of figure~\ref{fig:NS}. Indeed, within the allowed
region 
$\{\theta < 0 , \, \tau< \tau_3(\tan \theta) \}
\cup
\{\theta > 0, \, \tau < \tau_R( \tan \theta) \}$, 
the stability condition 
(\ref{E:st_cond}) fails to apply in the region 
$\{\theta<0;\,\tau< \tau_t(\tan \theta)\}\cup \{\theta>0;
\,\tau> \tau_t(\tan \theta)\}$, where $p<0$, and in a 
peculiarly shaped region of the positive $\theta$ quadrant, 
where $s^2-4p<0$. Since $s$ is positive everywhere outside 
these regions, it turns out that linear stability characterizes 
not only, as expected, the maxima, but also most of the saddle 
points at positive $\theta$.

\subsection{Central depleted well}
\label{S:CDW}
As discussed in section \ref{S:Dyn}, there is only one 
parameter-independent CDW fixed point: ${\mathbf x}= (1/\sqrt{2},0,-1/\sqrt 2)$, which, according to the signature of matrices $Q$ and $P$ (see \ref{E:mqp}), 
is always a saddle point. Indeed, no matter for the sign of 
$\Delta_2=-2\,(1+\nu)/\tau$, the quadratic form has always 
an undefined signature. Recalling that $\Delta_1=\Delta_3=0$, 
the coefficients involved  stability condition (\ref{E:st_cond}) 
have the quite simple form 
$ s= 4[\tau^2+(1+ \nu )^2]/{\tau^2}$ 
and
$ p= 4[\tau^2+4(1 + \nu )]/{\tau^2}$ yielding 
$\Delta = 16\, \left[2\,(\nu^2-1)\tau^2+(\nu+1)^4\right]/\tau^4$.
Hence, introducing the functions
\begin{equation}
\label{E:cdwB}
c_1(\nu) = 2\,\sqrt{-1-\nu}, 
\qquad c_2(\nu)=\sqrt{(1+\nu)^3/[2\,(1-\nu)]}
\end{equation}
the CDW saddle is actually stable in the parameter regions 
$\{\tau>c_1(\nu);\,\nu <- 1\}$, $\{\tau< c_2(\nu);\,|\nu| < 1\}$ 
and $\{\tau>0;\,\nu > 1\}$, and unstable elsewhere. Notice that 
the former zone can be also described as $\nu<\bar \nu_R(\tau)$, 
where $\nu_R$ is strictly related to the boundary of one of the 
forbidden zones of the NFP stability diagram. Actually, as we 
already noticed above, the points lying on such (forbidden) 
boundary $\tau=\tau_R(\tan \theta)$ have a CDW form, since 
$R({\alpha};\,\tau)=0$ there (see (\ref{E:forb}) and the
defining formulas for $X_j=X_j(R, \theta)$, $j=\, 1,2,3$ in section
\ref{S:NR}).
The stability diagram for the CDW fixed point is displayed in 
figure~\ref{fig:CDW}. Notice that, unlike figures~\ref{fig:DS} and \ref{fig:NS}, 
it is in the  plane of the parameters, $\tau$ and $\nu$. This is 
possible because, as we already remarked, there is always a 
single, parameter-independent CDW configuration.

\end{appendix}

\section*{References}

\end{document}